\documentclass[aps,prl,preprint,groupedaddress]{revtex4-1}

\usepackage{graphicx}
\usepackage{multirow}
\usepackage{tabularx}
\usepackage{array}
\usepackage{upgreek}
\usepackage{amsmath}
\usepackage{amssymb}
\usepackage{siunitx}
\usepackage[yyyymmdd,hhmmss]{datetime}

\usepackage{xcolor}
\usepackage{dcolumn}
\usepackage[final]{hyperref}

\definecolor{link}{rgb}{0,0,0.9}
\definecolor{cite}{rgb}{0,0,0.9}
\definecolor{url}{rgb}{0,0,0.9}
\hypersetup
{
    colorlinks=true,
    linkcolor=link,
    citecolor=cite,
    urlcolor=url,
}

\newcommand{\VTPI}{V_\text{TPI}}
\newcommand{\gtwoo}{g^{(2)}(0)}
\newcommand{\gtwot}{g^{(2)}(\tau)}

\begin{document}


\title{Overcoming correlation fluctuations in two-photon interference experiments with differently bright and independently blinking remote quantum emitters}

\author{Jonas H. Weber}
\email[]{j.weber@ihfg.uni-stuttgart.de}
\author{Jan Kettler}
\author{H\"useyin Vural}
\author{Markus M\"uller}
\author{Julian Maisch}
\author{Michael Jetter}
\author{Simone L. Portalupi}
\author{Peter Michler}
\email[]{p.michler@ihfg.uni-stuttgart.de}
\affiliation{Institut f\"ur Halbleiteroptik und Funktionelle Grenzfl\"achen, Center for Integrated Quantum Science and Technology (IQ{$^{ST}$}) and SCoPE, University of Stuttgart, Allmandring 3, 70569 Stuttgart, Germany}
\homepage[]{www.ihfg.physik.uni-stuttgart.de}


\begin{abstract}
	As a fundamental building block for quantum computation and communication protocols, the correct verification of the two-photon interference (TPI) contrast between two independent quantum light sources is of utmost importance. Here, we experimentally demonstrate how frequently present blinking dynamics and changes in emitter brightness critically affect the Hong-Ou-Mandel-type (HOM) correlation histograms of remote TPI experiments measured via the commonly utilized setup configuration. We further exploit this qualitative and quantitative explanation of the observed correlation dynamics to establish an alternative interferometer configuration, which is overcoming the discussed temporal fluctuations, giving rise to an error-free determination of the remote TPI visibility. We prove full knowledge of the obtained correlation by reproducing the measured correlation statistics via Monte-Carlo simulations. As exemplary system, we make use of two pairs of remote semiconductor quantum dots, however, the same conclusions apply for TPI experiments with flying qubits from any kind of remote solid state quantum emitters.
\end{abstract}



\maketitle

	\begin{figure*}
		\centering
		\includegraphics[width=0.9\linewidth]{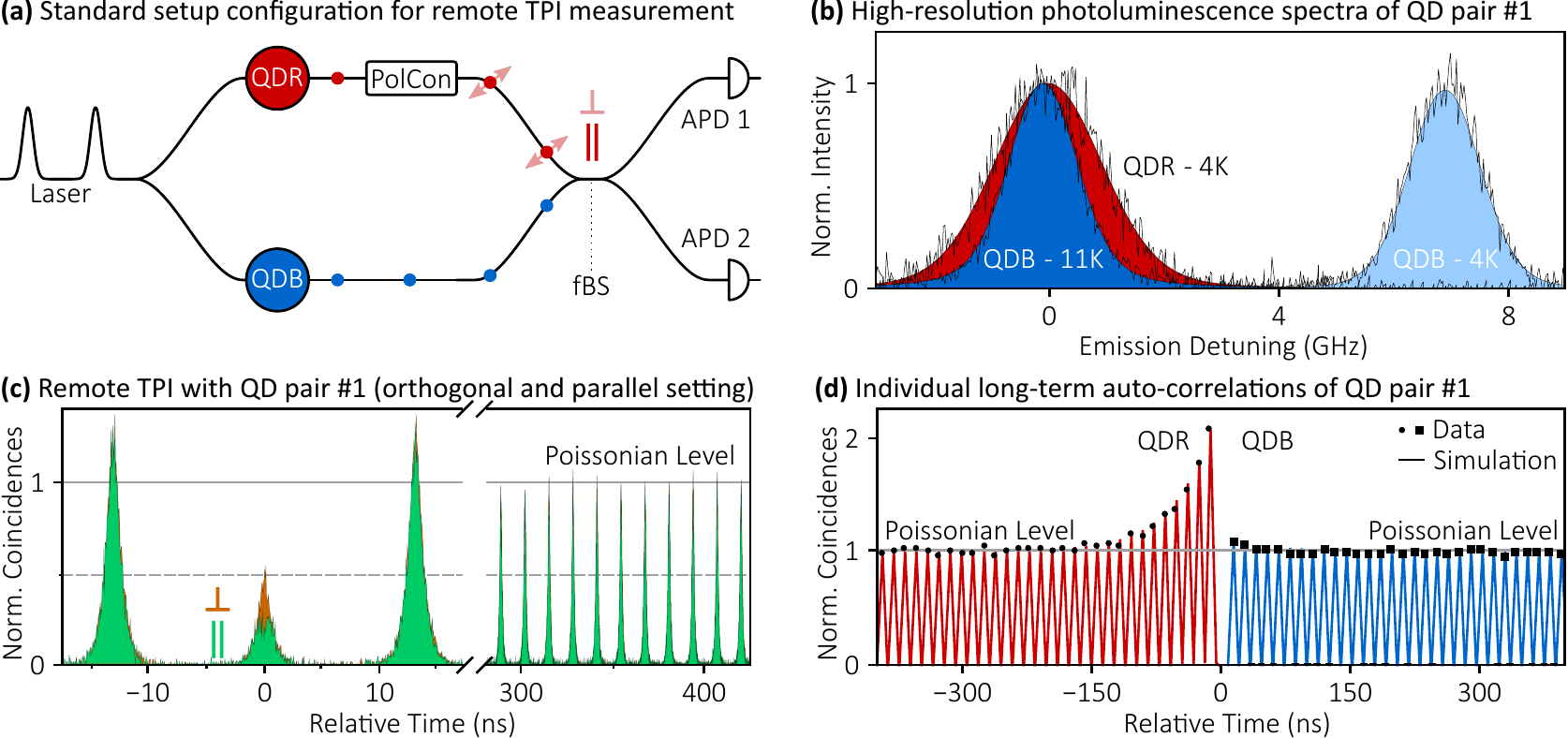}
		\caption{(a) Setup sketch for remote TPI measurements: polarization control allows for indistinguishable (parallel) and distinguishable (orthogonal) configuration with photon overlap on a fiber-based BS. (b) High-resolution PL spectra of QDR and QDB at different operation temperatures. (c) Remote TPI experiment with clear signature of quantum interference for parallel polarization setting. (d) Blinking of QDR is quantified via long-term auto-correlation measurement, while not observable for QDB. Solid lines represent Monte-Carlo simulations considering a nearby carrier trap (see Suppl. Info.).\label{fig:fig1}}
	\end{figure*}
Two-photon interference of single photons on a beam splitter is the fundamental building block in photonic based quantum technologies, such as boson sampling \cite{Wang2017,Loredo2016}, quantum repeaters \cite{Sangouard2007,Kimble2008} and quantum computing \cite{Knill2001}. High TPI visibilities require high photon indistinguishability of the interfering photons, i.e., the photons must be identical in all their optical properties, e.g., frequency, bandwidth and polarization \cite{Hong1987}. For many decades, spontaneous parametric down-conversion (SPDC) has been used as a source of single and entangled photons in quantum information experiments \cite{Ma2011,Pan2012,Takeoka2014}, benefiting from their high photon indistinguishability. In contrast to deterministic emitters, however, SPDC-sources suffer from Poisson statistics coming from the probabilistic photon generation. 

In recent years, solid-state quantum emitters, such as quantum dots (QD) and various color centers get increasing attention since they are capable to emit single photons on-demand \cite{Michler2000,Lounis2000,Santori2001,Gaebel2004,Wang2006,Hogele2008,Babinec2010} and can additionally be integrated into photonic chips. Encouraging results from several independent groups showed that single-photon sources based on QDs are especially promising candidates due to their high brightness and near-unity degree of indistinguishability of consecutively emitted photons \cite{He2013,Somaschi2016} exhibiting Fourier-transform limitation \cite{Kuhlmann2015}. For scaled quantum networks, however, TPI with photons from remote quantum emitters is crucial \cite{Riedmatten2003,Beugnon2006,Maunz2007,Patel2010,Lettow2010,Bernien2012}. Here, two major differences exist, when comparing TPI with consecutively emitted photons from an individual emitter and TPI with remote sources: first of all, photons from two distinct emitters are spectrally uncorrelated \cite{Legero2003,Gold2014,Wang2016}, i.e., broadening mechanisms due to the fluctuating solid state environment temporally evolve to their full extent. As a consequence, two emitters of individually very high TPI visibilities can than lead to drastically reduced remote HOM visibility (compare Table \ref{tab:tab1}). Secondly, temporal correlation within the individual photon streams, e.g., through so-called blinking of the emission \cite{Xie1994,Nirmal1996,Mason1998,Shimizu2001,Santori2001}, may result in unexpected photon statistics in the remote TPI experiment \cite{Reindl2017,Jons2017}. Commonly the TPI visibility $\VTPI$, is extracted via the definition \cite{Santori2002}
\begin{equation}
	\VTPI ^{(1)} =1-\frac{\gtwoo_\parallel}{\gtwoo_\perp}\label{eq:VTPI},
\end{equation}
where the auto-correlation function at zero-delay in presence of quantum interference is compared to the one without quantum interference, represented by $\gtwoo_\parallel$ and $\gtwoo_\perp$, respectively. However, a temporally changing normalization level within the respective correlation function $\gtwot$, due to a variation in blinking behavior or emitter brightness, leads to inconsistent normalization between $\gtwoo_\parallel$ and $\gtwoo_\perp$. As a consequence, the reported interpretations of the correlation statistics may strongly overestimate the actual photon indistinguishability, therefore, being often not in agreement with the theoretical expectation.
Here, we experimentally and theoretically investigate the influence of blinking and count rate fluctuations on the correlation statistics obtained with the commonly utilized setup configuration. We further show to which extent the Poissonian Level can be utilized to extract $\VTPI$ while having unbalanced photon flux. However, in some case the Poissonian Level can not be clearly identified, i.e., as blinking time scales can span multiple time scales \cite{Davanco2014}. Additionally, in on-chip applications, the necessary modification to monitor the sources photon flux are impractical to be implemented. As a consequence of the new insights, we present a novel setup configuration where we prove intrinsic stability of the coincidence statistics versus any kind of temporal correlation. The whole discussion is exemplified for two differently bright and independently blinking semiconductor QDs, while all correlation measurements are very well supported via Monte Carlo simulations \cite{Kettler2016PhD}.
	\begin{table}
		\centering
		\setlength{\extrarowheight}{2.5pt}
		\begin{tabular}{l l | c c c c c c }
			\hline
			\hline
			& &$V_\text{TPI,exp}^\text{single}$& $\tau_\text{dec}$& $\Delta\nu_\text{inh}$& $V_\text{TPI,sim}^\text{remote}$& $V_\text{TPI,exp}^\text{remote}$\\
			& &(\%) &(ps) &(GHz) &(\%) &(\%)\\ \hline			
			\multirow{2}{*}{\#1} &	QDR	&\num{72+-4}	& \num{580+-10}	& \num{2.0+-0.1}& \multirow{2}{*}{\num{27+-1}}&\multirow{2}{*}{\num{26+-3}}\\
			&QDB						&\num{58+-4}		& \num{600+-10}	& \num{1.3+-0.1}&	&	&\\ 
			\hline
			\multirow{2}{*}{\#2} &	QD1	&\num{96+-1}& \num{418+-12}	& \enspace\num{1.5+-0.5}$^\star$&	\multirow{2}{*}{\num{35+-1}}&\multirow{2}{*}{\num{42+-11}}\\
			&QD2							&\num{92+-1}& \num{509+-15}	& \num{1.3+-0.1}&	&	&\\ 
			\hline
			\hline
		\end{tabular}
		\caption{Characteristic parameter of QD-pair \#1 and \#2$^{\star\star}$: TPI visibility of consecutively emitted photons (\SI{4}{\nano\second} time delay), decay time and inhomogeneous linewidth. The latter two are utilized to simulate the remote TPI visibility following \cite{Legero2003,Vural2018}. Corresponding experimental remote TPI visibilities are reported. ($^\star$additional contribution from secondary line, see Suppl. Info.; $^{\star\star}$QD growth: \#1 \cite{Portalupi2016}; \#2 \cite{He2013})}
		\label{tab:tab1}
	\end{table}

In this study, we present the results of two remote TPI experiments carried out with two different QD-pairs. Each pair consists of two distinct QDs further denoted as QDR/QDB and QD1/QD2 being situated in distinct cryostats (see Figure \ref{fig:fig1}a). We deterministically initialize the charged exciton transition of the QDs, via a resonant and coherent $\pi$-pulse excitation scheme. After state initialization, the resonance fluorescence is filtered via confocal polarization suppression of the residual laser light. A monochromator is further utilized to spectrally remove the phonon-sidebands. Before TPI is carried out using a fiber-based beam splitter (fBS), quarter- and halfwave plates are utilized to control the relative polarization state (PolCon) of the two emitter arms.
\begin{figure*}
	\centering
	\includegraphics[width=0.9\linewidth]{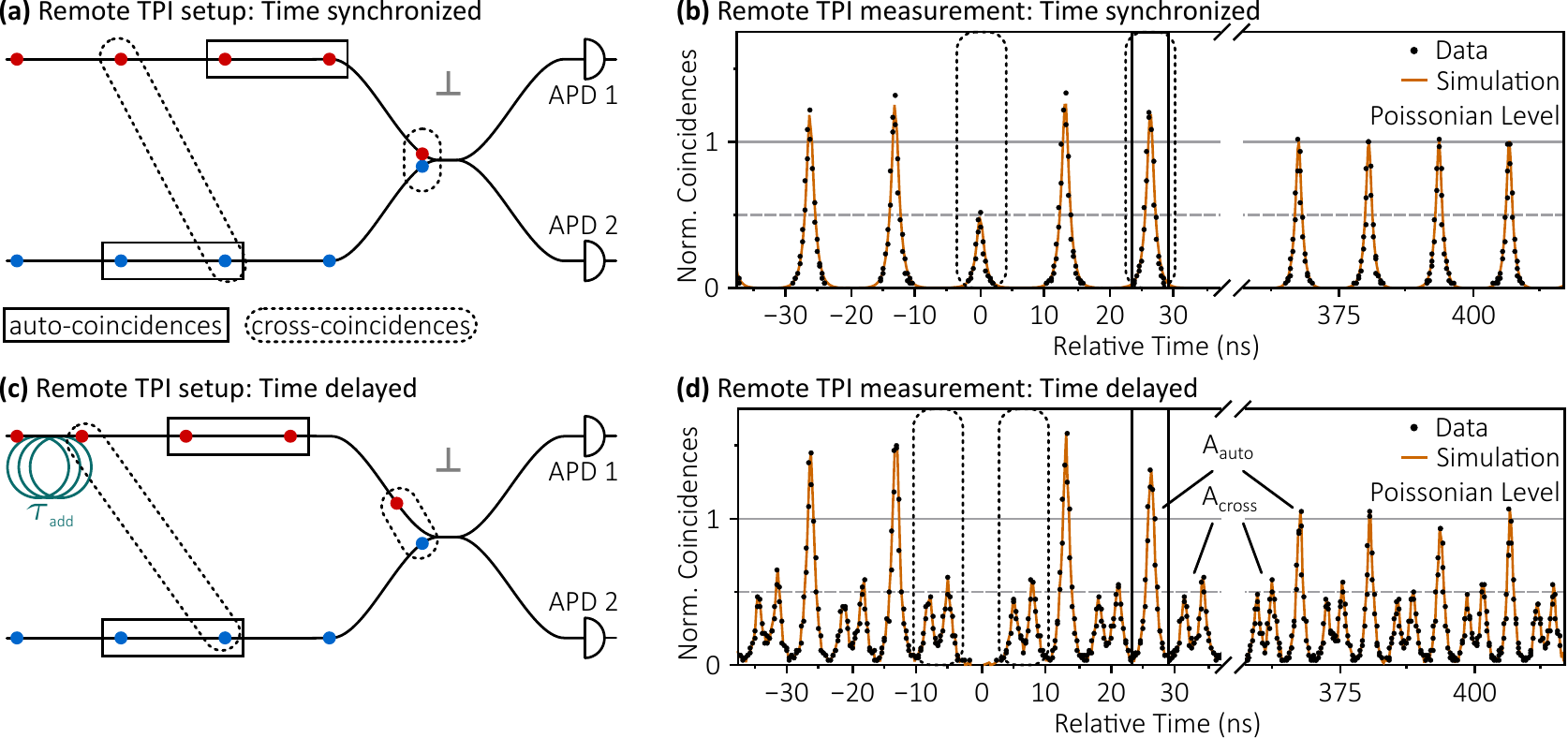}
	\caption{(a) Exemplification of the two types of coincidences: `auto-coincidences' $A_\text{auto}$ (solid boxes) corresponding to events within the photon stream of an individual emitter, `cross-coincidences' $A_\text{cross}$ (dashed boxes) corresponding to events between the two emitters. (b) Resulting remote TPI measurement with QDR and QDB for orthogonal polarization setting with respective data from simulation. (c) Breaking of time synchronization of the two photon streams via a 1-m fiber ($t_\text{add}=\SI{5}{\nano\second}$). (d) Resulting TPI measurement exhibiting separation of cross- from auto-coincidences.\label{fig:fig2}}
\end{figure*}
In both cases, thermal tuning is applied to tune the respectively selected transitions into spectral resonance, as it is exemplified for QD pair \#1 in Figure \ref{fig:fig1}b. The resulting remote TPI measurement shows clear signature of quantum interference (Figure \ref{fig:fig1}c and Suppl. Info.), indicated by the absence of coincidence counts in the zero delay peak of the parallel polarization setting. Even though starting with two different TPI levels for the individual emitters, the resulting remote TPI contrast of QD-pair \#1 and \#2 are on the same order as it is shown in Table \ref{tab:tab1}. This is based on the fact that the radiative lifetime as well as inhomogeneous broadening $\Delta\nu_\text{inh}$, due to spin and charge noise \cite{Robinson2000,Kuhlmann2015}, fully determine the TPI visibility of the remote emitters \cite{Legero2003}. Here, measurements of decay time $\tau_\text{dec}$ and high-resolution PL (hPL) lead to very good agreement between simulation and experiment (see Table \ref{tab:tab1}). In practice, however, the TPI visibility is mostly determined by measuring the photon coalescence on a beam splitter. In the following, we will discuss how to overcome any kind of correlation fluctuations, exemplified via QD-pair \#1, to determine the TPI visibility such that it is in agreement with the theoretical expectation.   

In the analysis of remote TPI data, utilizing the standard setup configuration, it is common to normalize the obtained $\gtwot$ on the first repetitions of the pulsed correlation measurement \cite{Gold2014,Giesz2015,Thoma2017,Reindl2017,Zopf2017}. Working with single photons, one would expect the distinguishable case to be $\gtwoo _{\perp,\text{theo}}=0.5$ \cite{Hong1987}. However, a comparison of the first side and the zero-delay peak in Figure \ref{fig:fig1}c reveals $\gtwoo_{\perp,\text{exp,first}} = \num{0.371 +- 0.013}$. The origin of this mismatch can be identified via long-term auto-correlation measurements of the two QDs (Figure \ref{fig:fig1}d): while QDB exhibits little deviation from the Poissonian level, QDR shows a clear signature of clustering in the single-photon stream with a bunching time constant on the order of $\SI{100}{\nano\second}$. The underlying effect can be attributed to the charge environment of the QD, where the ability to resonantly excite a distinct QD transition, can be switched on and off according to a randomly varying charge state (compare Suppl. Info. and \cite{Nguyen2013,Kettler2016PhD}).

\begin{figure*}
	\centering
	\includegraphics[width=0.9\linewidth]{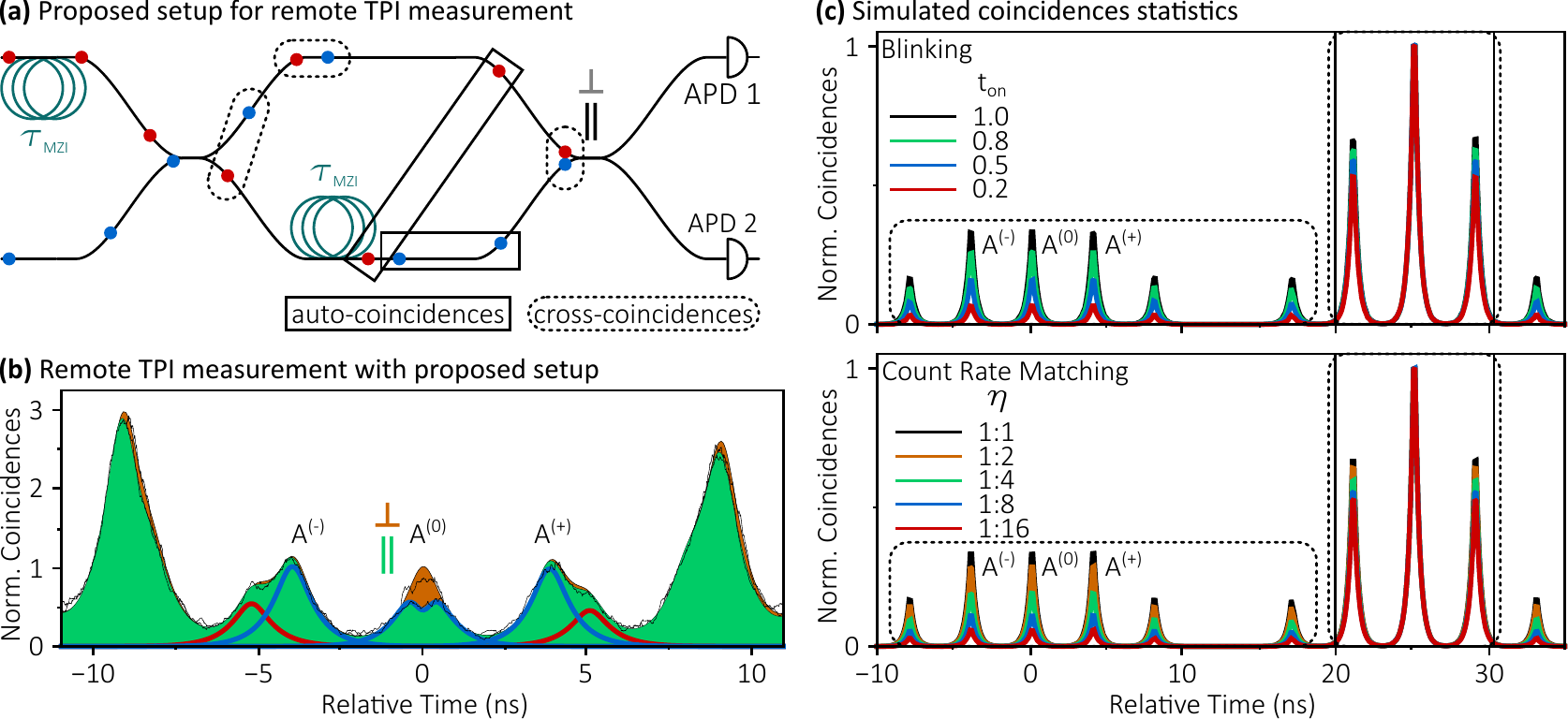}
	\caption{(a) Unbalanced MZI setup enabling self-normalization on cross-coincidences, independently of blinking dynamics and brightness fluctuations. (b) Remote TPI measurement with QDR and QDB following the sketched approach: despite a count rate ratio $\eta=0.5$ and aforementioned blinking dynamics, a $\gtwoo_\text{$\perp$,exp} \approx 0.5$ is found in accordance to the expectation. (c)~Prove of constant ratio of center peak triplet ($A^{(-)}$, $A^{(0)}$, $A^{(+)}$), in presence of blinking (top) or count rate mismatch (bottom), then enabling the mentioned self-normalization.  \label{fig:fig3}}
\end{figure*}

It is therefore necessary, to get further insight in the possible coincidence contribution to the observed bunching signature. Interestingly, a slight modification of the standard setup configuration (Figure \ref{fig:fig2}a) via additional 1\,m-fiber-delay (corresponding to $\tau_\text{add}\approx\SI{5}{\nano\second}$) allows to identify two different types of contributions: `auto-coincidences' and `cross-coincidences'. Auto coincidences $A_\text{auto}$ (solid boxes) occur only within the same photon stream while cross-coincidences $A_\text{cross}$ (dashed boxes) only occur between the two photon streams. Figure \ref{fig:fig2}b shows a correlation measurement resulting from the standard setup, where the auto-coincidences are expected to accumulate at the laser repetitions (at $\pm m\cdot \tau_\text{rep}$, with $m\in\mathbb{N}^+$) while cross-coincidences additionally accumulate the zero-delay peak. Via the slight modification (Figure \ref{fig:fig2}c), the difference between auto- and cross-coincidences becomes clear when comparing the resulting correlation statistics in Figure \ref{fig:fig2}d. Cross-coincidences are now shifted by $\pm \tau_\text{add}$ from the laser repetitions and zero-time delay, whereas auto-coincidences stay at the repetitions. From the correlation it appears that the blinking behavior can only be attributed to auto-coincidences. In contrast, cross-coincidences do not show any kind of long-term correlation - they are obviously insensitive to blinking and count rate fluctuations. Consequently, even normalization on the Poissonian level can be inconsistent between different measurements as blinking behavior and photon fluctuations change over time. Hence, comparison between two separate measurements should be avoided. Therefore, the previous qualitative discussion is utilized to extrapolate instead of directly measuring the orthogonal case. 

As discussed, the zero-delay peak $A_0$, which only includes cross-coincidences, has to be normalized to the side peaks $A_\text{side}$, i.e., the Poissonian level including both auto- and cross-coincidences. This results in $\gtwoo_\text{$\perp$,theo} = A_0/A_\text{side} = A_\text{cross}/\left(A_\text{auto}+A_\text{cross}\right).$ Next, matching of count rates $c_{1,2}$ of the photon streams is introduced as $\eta\equiv c_2/c_1$ (with $c_2<c_1$). In case of perfect beam splitting ratio, this leads then to $\gtwoo_{\perp,\text{extra}}=2\eta/(\eta+1)^2$ (compare Suppl. Info.). This expression can be utilize to extrapolate $\gtwoo_\perp$ in dependences of emitter count rates. Following the common procedure, $\VTPI$ can then be calculated by comparison of the experimental $\gtwoo_\text{$\parallel$,exp}$ with the extrapolated $\gtwoo_\text{$\perp$,extra}$ -- two values obtained from a single measurement -- leading to
\begin{equation}
	\VTPI ^{(2)} =1-\frac{\gtwoo_{\parallel,\text{exp}}}{\gtwoo_{\perp,\text{extra}}}\label{eq:VTPI2}.
\end{equation}
As long as emitter count rates are monitored over the whole measurement, the TPI contrast can effectively be extracted from a single measurement even with differently bright and blinking emitters. In case of \ref{fig:fig1}c, we extract $\gtwoo_\text{$\parallel$,exp} = \num{0.370+-0.022}$ and extrapolate $\gtwoo_\text{$\perp$,extra} = \num{0.499+-0.004}$ from the detector time traces. According to (\ref{eq:VTPI2}) this results in $\VTPI ^{(2)} = \num{25.7+-4.9}$, and $\VTPI ^{(2)} = \num{26+-5}$ after beam splitter correction.

Even though we demonstrated how to extract $\VTPI$ with the standard setup configuration and additional monitoring of the photon flux, relying on the Poissonian level is not always possible (see introduction). However, the investigations emphasize that cross-coincidences are unaffected by the mentioned correlation fluctuations. As a consequence, here, we present a novel setup configuration where normalization relies on cross-coincidences only. This implies an intrinsic stability to any kind of fluctuation of blinking or emitter brightness. Consequently, systematic errors are avoided, hence, reduced to pure statistical errors, while allowing for maximum photon flux even without the need for additional count rate monitoring. 

Figure \ref{fig:fig3}a shows an unbalanced MZI with delay $\tau_\text{MZI}$, commonly utilized to determine the indistinguishability of consecutively emitted photons from an individual QD. In this case, however, it is fed by the two distinct emitters whose photon stream is initially unsynchronized by $\tau_\text{MZI}$. First of all, at zero time delay cross-coincidences are synchronized in time, enabling TPI. Secondly, parts of the cross-coincidences which are unsynchronized ($\tau\neq 0$) are still separated from auto-coincidences (e.g., at $\tau = \pm
\tau_\text{MZI}$). As discussed, cross-coincidences are unaffected by blinking or count rate mismatch: this reflects in a constant central peaks ratio, allowing for self-normalization. 

The experimental result is shown in Figure \ref{fig:fig3}b with clear signature of TPI at zero time delay in the case of parallel polarization setting. Again $\VTPI$ is calculated via (\ref{eq:VTPI}), where the individual polarization setting is now determined by normalization of the zero-delay peak $A^{(0)}$ to the first side-peaks $A^{(-,+)}$ (no Poissonian level is necessary). In the presented case, the first side-peaks are already merging with coincidences of the first repetition, still the cumulative peak consist only of cross-coincidences. The measurement was performed with large count rate mismatch of ($\eta=0.5$) and under the same blinking dynamics as shown in Figure \ref{fig:fig1}d. Nevertheless, the distinguishable case shows $\gtwoo_\text{$\perp$,exp} = \num{0.506+-0.029}$ being in accordance to the prediction. i.e., $\gtwoo_\text{$\perp$,extra} = \num{0.506+-0.002}$ (deviation from 0.5 due to beam splitter ratios \cite{Santori2002}). To further strengthen the experimental verification, that the ratio of the center peak triplet ($A^{(-)}$, $A^{(0)}$, $A^{(+)}$) is unaffected, Figure \ref{fig:fig3}c shows simulations of the center quintuplet and the quintuplet at $\tau_\text{rep}$. Here, the repetition is chosen to be separated from the center, preventing time overlap. As the blinking ratio $t_\text{on}$ is decreased (top), i.e., the off time is increased, hence bunching more pronounced (compare \cite{Santori2001}). Peaks with a mixture of auto- and cross-events, then, change peak ratios. The peak ratio of the center quintuplet, however, is unaffected. The same is true for varying count rate matching $\eta$ of the two emitters (bottom). If this setup scheme is utilized, the TPI contrast can be extracted from a single measurement, by applying (\ref{eq:VTPI}) and extrapolation of the cross-polarized correlation $\gtwoo_\text{$\perp$,extra}$, leading to 
\begin{equation}
	\VTPI^{(3)}= 1-\frac{\gtwoo_\text{$\parallel$,exp}}{\gtwoo_\text{$\perp$,extra}} = 1-\frac{A^{(0)}/\left(A^{(-)}+A^{(+)}\right)}{\gtwoo_\text{$\perp$,extra}}\label{eq:VTPI3}.
\end{equation}
Considering peak overlap, the measurement in Figure \ref{fig:fig3}b reveals $\gtwoo_\text{$\parallel$,exp} = \num{0.379+-0.013}$. Following (\ref{eq:VTPI3}), this leads to $\VTPI ^{(3)} = \num{25.2+-2.8}$, and $\VTPI ^{(2)} = \num{26+-3}$ after beam splitter correction, being in accordance to the standard approach shown in Figure \ref{fig:fig1}c.

In this study, we demonstrated how blinking dynamics as well as the sources individual count rate influences determination the coincidence contribution in remote TPI experiments. This allows for the correct estimation of $\VTPI$ with single measurements, by means of both monitoring count rates and isolating the short-term correlations driven by the blinking dynamics of the quantum emitters. Furthermore, we identify two types of coincidences: auto-coincidences, which are affected by temporal correlations and cross-coincidences, which are intrinsically stable. As a result, we introduce a new method to measure the indistinguishability of photons from remote emitters where normalization only relies on stable cross-coincidences. This method becomes particularly relevant not only for the practical experimental implementation but most importantly because it allows the precise and correct evaluation of the remote sources interference without being affected by unavoidable correlation dynamics. This is of fundamental interest for practical implementation of quantum technology, such as Boson sampling, where the reliability and reproducibility of such building block operation is of key importance. 

\noindent We exemplary conducted remote TPI using semiconductor QDs, taking advantage of their superior brightness and state-of-the-art performances. However, the same conclusions and methods can be applied to any pair of quantum emitter which can be used to perform two-photon interference. This study helps in clearly understanding remote TPI which will be of key importance for future developments of quantum technology.

\begin{acknowledgments}
We thank C. Schneider and S. H\"ofling for providing us with the high quality MBE-sample including QD-pair \#2. Furthermore, we thank S. Kern for technical assistance and acknowledge financial support by the DFG via the project MI 500/27-1.
\end{acknowledgments}

\begin{thebibliography}{44}%
	\makeatletter
	\providecommand \@ifxundefined [1]{%
		\@ifx{#1\undefined}
	}%
	\providecommand \@ifnum [1]{%
		\ifnum #1\expandafter \@firstoftwo
		\else \expandafter \@secondoftwo
		\fi
	}%
	\providecommand \@ifx [1]{%
		\ifx #1\expandafter \@firstoftwo
		\else \expandafter \@secondoftwo
		\fi
	}%
	\providecommand \natexlab [1]{#1}%
	\providecommand \enquote  [1]{``#1''}%
	\providecommand \bibnamefont  [1]{#1}%
	\providecommand \bibfnamefont [1]{#1}%
	\providecommand \citenamefont [1]{#1}%
	\providecommand \href@noop [0]{\@secondoftwo}%
	\providecommand \href [0]{\begingroup \@sanitize@url \@href}%
	\providecommand \@href[1]{\@@startlink{#1}\@@href}%
	\providecommand \@@href[1]{\endgroup#1\@@endlink}%
	\providecommand \@sanitize@url [0]{\catcode `\\12\catcode `\$12\catcode
		`\&12\catcode `\#12\catcode `\^12\catcode `\_12\catcode `\%12\relax}%
	\providecommand \@@startlink[1]{}%
	\providecommand \@@endlink[0]{}%
	\providecommand \url  [0]{\begingroup\@sanitize@url \@url }%
	\providecommand \@url [1]{\endgroup\@href {#1}{\urlprefix }}%
	\providecommand \urlprefix  [0]{URL }%
	\providecommand \Eprint [0]{\href }%
	\providecommand \doibase [0]{http://dx.doi.org/}%
	\providecommand \selectlanguage [0]{\@gobble}%
	\providecommand \bibinfo  [0]{\@secondoftwo}%
	\providecommand \bibfield  [0]{\@secondoftwo}%
	\providecommand \translation [1]{[#1]}%
	\providecommand \BibitemOpen [0]{}%
	\providecommand \bibitemStop [0]{}%
	\providecommand \bibitemNoStop [0]{.\EOS\space}%
	\providecommand \EOS [0]{\spacefactor3000\relax}%
	\providecommand \BibitemShut  [1]{\csname bibitem#1\endcsname}%
	\let\auto@bib@innerbib\@empty
	\bibitem [{\citenamefont {Wang}\ \emph {et~al.}(2017)\citenamefont {Wang},
		\citenamefont {He}, \citenamefont {Li}, \citenamefont {Su}, \citenamefont
		{Li}, \citenamefont {Huang}, \citenamefont {Ding}, \citenamefont {Chen},
		\citenamefont {Liu}, \citenamefont {Qin}, \citenamefont {Li}, \citenamefont
		{He}, \citenamefont {Schneider}, \citenamefont {Kamp}, \citenamefont {Peng},
		\citenamefont {H{\"{o}}fling}, \citenamefont {Lu},\ and\ \citenamefont
		{Pan}}]{Wang2017}%
	\BibitemOpen
	\bibfield  {author} {\bibinfo {author} {\bibfnamefont {H.}~\bibnamefont
			{Wang}}, \bibinfo {author} {\bibfnamefont {Y.}~\bibnamefont {He}}, \bibinfo
		{author} {\bibfnamefont {Y.-H.}\ \bibnamefont {Li}}, \bibinfo {author}
		{\bibfnamefont {Z.-E.}\ \bibnamefont {Su}}, \bibinfo {author} {\bibfnamefont
			{B.}~\bibnamefont {Li}}, \bibinfo {author} {\bibfnamefont {H.-L.}\
			\bibnamefont {Huang}}, \bibinfo {author} {\bibfnamefont {X.}~\bibnamefont
			{Ding}}, \bibinfo {author} {\bibfnamefont {M.-C.}\ \bibnamefont {Chen}},
		\bibinfo {author} {\bibfnamefont {C.}~\bibnamefont {Liu}}, \bibinfo {author}
		{\bibfnamefont {J.}~\bibnamefont {Qin}}, \bibinfo {author} {\bibfnamefont
			{J.-P.}\ \bibnamefont {Li}}, \bibinfo {author} {\bibfnamefont {Y.-M.}\
			\bibnamefont {He}}, \bibinfo {author} {\bibfnamefont {C.}~\bibnamefont
			{Schneider}}, \bibinfo {author} {\bibfnamefont {M.}~\bibnamefont {Kamp}},
		\bibinfo {author} {\bibfnamefont {C.-Z.}\ \bibnamefont {Peng}}, \bibinfo
		{author} {\bibfnamefont {S.}~\bibnamefont {H{\"{o}}fling}}, \bibinfo {author}
		{\bibfnamefont {C.-Y.}\ \bibnamefont {Lu}}, \ and\ \bibinfo {author}
		{\bibfnamefont {J.-W.}\ \bibnamefont {Pan}},\ }\href {\doibase
		10.1038/nphoton.2017.63} {\bibfield  {journal} {\bibinfo  {journal} {Nat.
				Photonics}\ }\textbf {\bibinfo {volume} {11}},\ \bibinfo {pages} {361}
		(\bibinfo {year} {2017})}\BibitemShut {NoStop}%
	\bibitem [{\citenamefont {Loredo}\ \emph {et~al.}(2016)\citenamefont {Loredo},
		\citenamefont {Zakaria}, \citenamefont {Somaschi}, \citenamefont {Anton},
		\citenamefont {de~Santis}, \citenamefont {Giesz}, \citenamefont {Grange},
		\citenamefont {Broome}, \citenamefont {Gazzano}, \citenamefont {Coppola},
		\citenamefont {Sagnes}, \citenamefont {Lemaitre}, \citenamefont {Auffeves},
		\citenamefont {Senellart}, \citenamefont {Almeida},\ and\ \citenamefont
		{White}}]{Loredo2016}%
	\BibitemOpen
	\bibfield  {author} {\bibinfo {author} {\bibfnamefont {J.~C.}\ \bibnamefont
			{Loredo}}, \bibinfo {author} {\bibfnamefont {N.~A.}\ \bibnamefont {Zakaria}},
		\bibinfo {author} {\bibfnamefont {N.}~\bibnamefont {Somaschi}}, \bibinfo
		{author} {\bibfnamefont {C.}~\bibnamefont {Anton}}, \bibinfo {author}
		{\bibfnamefont {L.}~\bibnamefont {de~Santis}}, \bibinfo {author}
		{\bibfnamefont {V.}~\bibnamefont {Giesz}}, \bibinfo {author} {\bibfnamefont
			{T.}~\bibnamefont {Grange}}, \bibinfo {author} {\bibfnamefont {M.~A.}\
			\bibnamefont {Broome}}, \bibinfo {author} {\bibfnamefont {O.}~\bibnamefont
			{Gazzano}}, \bibinfo {author} {\bibfnamefont {G.}~\bibnamefont {Coppola}},
		\bibinfo {author} {\bibfnamefont {I.}~\bibnamefont {Sagnes}}, \bibinfo
		{author} {\bibfnamefont {A.}~\bibnamefont {Lemaitre}}, \bibinfo {author}
		{\bibfnamefont {A.}~\bibnamefont {Auffeves}}, \bibinfo {author}
		{\bibfnamefont {P.}~\bibnamefont {Senellart}}, \bibinfo {author}
		{\bibfnamefont {M.~P.}\ \bibnamefont {Almeida}}, \ and\ \bibinfo {author}
		{\bibfnamefont {A.~G.}\ \bibnamefont {White}},\ }\href {\doibase
		10.1364/OPTICA.3.000433} {\bibfield  {journal} {\bibinfo  {journal} {Optica}\
		}\textbf {\bibinfo {volume} {3}},\ \bibinfo {pages} {433} (\bibinfo {year}
		{2016})}\BibitemShut {NoStop}%
	\bibitem [{\citenamefont {Sangouard}\ \emph {et~al.}(2007)\citenamefont
		{Sangouard}, \citenamefont {Simon}, \citenamefont {Min{\'{a}}{\v{r}}},
		\citenamefont {Zbinden}, \citenamefont {de~Riedmatten},\ and\ \citenamefont
		{Gisin}}]{Sangouard2007}%
	\BibitemOpen
	\bibfield  {author} {\bibinfo {author} {\bibfnamefont {N.}~\bibnamefont
			{Sangouard}}, \bibinfo {author} {\bibfnamefont {C.}~\bibnamefont {Simon}},
		\bibinfo {author} {\bibfnamefont {J.}~\bibnamefont {Min{\'{a}}{\v{r}}}},
		\bibinfo {author} {\bibfnamefont {H.}~\bibnamefont {Zbinden}}, \bibinfo
		{author} {\bibfnamefont {H.}~\bibnamefont {de~Riedmatten}}, \ and\ \bibinfo
		{author} {\bibfnamefont {N.}~\bibnamefont {Gisin}},\ }\href {\doibase
		10.1103/PhysRevA.76.050301} {\bibfield  {journal} {\bibinfo  {journal} {Phys.
				Rev. A}\ }\textbf {\bibinfo {volume} {76}},\ \bibinfo {pages} {050301}
		(\bibinfo {year} {2007})}\BibitemShut {NoStop}%
	\bibitem [{\citenamefont {Kimble}(2008)}]{Kimble2008}%
	\BibitemOpen
	\bibfield  {author} {\bibinfo {author} {\bibfnamefont {H.~J.}\ \bibnamefont
			{Kimble}},\ }\href {\doibase 10.1038/nature07127} {\bibfield  {journal}
		{\bibinfo  {journal} {Nature}\ }\textbf {\bibinfo {volume} {453}},\ \bibinfo
		{pages} {1023} (\bibinfo {year} {2008})}\BibitemShut {NoStop}%
	\bibitem [{\citenamefont {Knill}\ \emph {et~al.}(2001)\citenamefont {Knill},
		\citenamefont {Laflamme},\ and\ \citenamefont {Milburn}}]{Knill2001}%
	\BibitemOpen
	\bibfield  {author} {\bibinfo {author} {\bibfnamefont {E.}~\bibnamefont
			{Knill}}, \bibinfo {author} {\bibfnamefont {R.}~\bibnamefont {Laflamme}}, \
		and\ \bibinfo {author} {\bibfnamefont {G.~J.}\ \bibnamefont {Milburn}},\
	}\href {\doibase 10.1038/35051009} {\bibfield  {journal} {\bibinfo  {journal}
			{Nature}\ }\textbf {\bibinfo {volume} {409}},\ \bibinfo {pages} {46}
		(\bibinfo {year} {2001})}\BibitemShut {NoStop}%
	\bibitem [{\citenamefont {Hong}\ \emph {et~al.}(1987)\citenamefont {Hong},
		\citenamefont {Ou},\ and\ \citenamefont {Mandel}}]{Hong1987}%
	\BibitemOpen
	\bibfield  {author} {\bibinfo {author} {\bibfnamefont {C.~K.}\ \bibnamefont
			{Hong}}, \bibinfo {author} {\bibfnamefont {Z.~Y.}\ \bibnamefont {Ou}}, \ and\
		\bibinfo {author} {\bibfnamefont {L.}~\bibnamefont {Mandel}},\ }\href
	{\doibase 10.1103/PhysRevLett.59.2044} {\bibfield  {journal} {\bibinfo
			{journal} {Phys. Rev. Lett.}\ }\textbf {\bibinfo {volume} {59}},\ \bibinfo
		{pages} {2044} (\bibinfo {year} {1987})}\BibitemShut {NoStop}%
	\bibitem [{\citenamefont {Ma}\ \emph {et~al.}(2011)\citenamefont {Ma},
		\citenamefont {Zotter}, \citenamefont {Kofler}, \citenamefont {Jennewein},\
		and\ \citenamefont {Zeilinger}}]{Ma2011}%
	\BibitemOpen
	\bibfield  {author} {\bibinfo {author} {\bibfnamefont {X.-s.}\ \bibnamefont
			{Ma}}, \bibinfo {author} {\bibfnamefont {S.}~\bibnamefont {Zotter}}, \bibinfo
		{author} {\bibfnamefont {J.}~\bibnamefont {Kofler}}, \bibinfo {author}
		{\bibfnamefont {T.}~\bibnamefont {Jennewein}}, \ and\ \bibinfo {author}
		{\bibfnamefont {A.}~\bibnamefont {Zeilinger}},\ }\href {\doibase
		10.1103/PhysRevA.83.043814} {\bibfield  {journal} {\bibinfo  {journal} {Phys.
				Rev. A}\ }\textbf {\bibinfo {volume} {83}},\ \bibinfo {pages} {043814}
		(\bibinfo {year} {2011})}\BibitemShut {NoStop}%
	\bibitem [{\citenamefont {Pan}\ \emph {et~al.}(2012)\citenamefont {Pan},
		\citenamefont {Chen}, \citenamefont {Lu}, \citenamefont {Weinfurter},
		\citenamefont {Zeilinger},\ and\ \citenamefont {{\.{Z}}ukowski}}]{Pan2012}%
	\BibitemOpen
	\bibfield  {author} {\bibinfo {author} {\bibfnamefont {J.-W.}\ \bibnamefont
			{Pan}}, \bibinfo {author} {\bibfnamefont {Z.-B.}\ \bibnamefont {Chen}},
		\bibinfo {author} {\bibfnamefont {C.-Y.}\ \bibnamefont {Lu}}, \bibinfo
		{author} {\bibfnamefont {H.}~\bibnamefont {Weinfurter}}, \bibinfo {author}
		{\bibfnamefont {A.}~\bibnamefont {Zeilinger}}, \ and\ \bibinfo {author}
		{\bibfnamefont {M.}~\bibnamefont {{\.{Z}}ukowski}},\ }\href {\doibase
		10.1103/RevModPhys.84.777} {\bibfield  {journal} {\bibinfo  {journal} {Rev.
				Mod. Phys.}\ }\textbf {\bibinfo {volume} {84}},\ \bibinfo {pages} {777}
		(\bibinfo {year} {2012})}\BibitemShut {NoStop}%
	\bibitem [{\citenamefont {Takeoka}\ \emph {et~al.}(2015)\citenamefont
		{Takeoka}, \citenamefont {Jin},\ and\ \citenamefont {Sasaki}}]{Takeoka2014}%
	\BibitemOpen
	\bibfield  {author} {\bibinfo {author} {\bibfnamefont {M.}~\bibnamefont
			{Takeoka}}, \bibinfo {author} {\bibfnamefont {R.-B.}\ \bibnamefont {Jin}}, \
		and\ \bibinfo {author} {\bibfnamefont {M.}~\bibnamefont {Sasaki}},\ }\href
	{\doibase 10.1088/1367-2630/17/4/043030} {\bibfield  {journal} {\bibinfo
			{journal} {New J. Phys.}\ }\textbf {\bibinfo {volume} {17}},\ \bibinfo
		{pages} {043030} (\bibinfo {year} {2015})}\BibitemShut {NoStop}%
	\bibitem [{\citenamefont {Michler}\ \emph {et~al.}(2000)\citenamefont
		{Michler}, \citenamefont {Kiraz}, \citenamefont {Becher}, \citenamefont
		{Schoenfeld}, \citenamefont {Petroff}, \citenamefont {Zhang}, \citenamefont
		{Hu},\ and\ \citenamefont {{\. I}mamo{\u g}lu}}]{Michler2000}%
	\BibitemOpen
	\bibfield  {author} {\bibinfo {author} {\bibfnamefont {P.}~\bibnamefont
			{Michler}}, \bibinfo {author} {\bibfnamefont {A.}~\bibnamefont {Kiraz}},
		\bibinfo {author} {\bibfnamefont {C.}~\bibnamefont {Becher}}, \bibinfo
		{author} {\bibfnamefont {W.~V.}\ \bibnamefont {Schoenfeld}}, \bibinfo
		{author} {\bibfnamefont {P.~M.}\ \bibnamefont {Petroff}}, \bibinfo {author}
		{\bibfnamefont {L.}~\bibnamefont {Zhang}}, \bibinfo {author} {\bibfnamefont
			{E.}~\bibnamefont {Hu}}, \ and\ \bibinfo {author} {\bibfnamefont
			{A.}~\bibnamefont {{\. I}mamo{\u g}lu}},\ }\href {\doibase
		10.1126/science.290.5500.2282} {\bibfield  {journal} {\bibinfo  {journal}
			{Science}\ }\textbf {\bibinfo {volume} {290}},\ \bibinfo {pages} {2282}
		(\bibinfo {year} {2000})}\BibitemShut {NoStop}%
	\bibitem [{\citenamefont {Lounis}\ and\ \citenamefont
		{Moerner}(2000)}]{Lounis2000}%
	\BibitemOpen
	\bibfield  {author} {\bibinfo {author} {\bibfnamefont {B.}~\bibnamefont
			{Lounis}}\ and\ \bibinfo {author} {\bibfnamefont {W.~E.}\ \bibnamefont
			{Moerner}},\ }\href {\doibase 10.1038/35035032} {\bibfield  {journal}
		{\bibinfo  {journal} {Nature}\ }\textbf {\bibinfo {volume} {407}},\ \bibinfo
		{pages} {491} (\bibinfo {year} {2000})}\BibitemShut {NoStop}%
	\bibitem [{\citenamefont {Santori}\ \emph {et~al.}(2001)\citenamefont
		{Santori}, \citenamefont {Pelton}, \citenamefont {Solomon}, \citenamefont
		{Dale},\ and\ \citenamefont {Yamamoto}}]{Santori2001}%
	\BibitemOpen
	\bibfield  {author} {\bibinfo {author} {\bibfnamefont {C.}~\bibnamefont
			{Santori}}, \bibinfo {author} {\bibfnamefont {M.}~\bibnamefont {Pelton}},
		\bibinfo {author} {\bibfnamefont {G.}~\bibnamefont {Solomon}}, \bibinfo
		{author} {\bibfnamefont {Y.}~\bibnamefont {Dale}}, \ and\ \bibinfo {author}
		{\bibfnamefont {Y.}~\bibnamefont {Yamamoto}},\ }\href {\doibase
		10.1103/PhysRevLett.86.1502} {\bibfield  {journal} {\bibinfo  {journal}
			{Phys. Rev. Lett.}\ }\textbf {\bibinfo {volume} {86}},\ \bibinfo {pages}
		{1502} (\bibinfo {year} {2001})}\BibitemShut {NoStop}%
	\bibitem [{\citenamefont {Gaebel}\ \emph {et~al.}(2004)\citenamefont {Gaebel},
		\citenamefont {Popa}, \citenamefont {Gruber}, \citenamefont {Domhan},
		\citenamefont {Jelezko},\ and\ \citenamefont {Wrachtrup}}]{Gaebel2004}%
	\BibitemOpen
	\bibfield  {author} {\bibinfo {author} {\bibfnamefont {T.}~\bibnamefont
			{Gaebel}}, \bibinfo {author} {\bibfnamefont {I.}~\bibnamefont {Popa}},
		\bibinfo {author} {\bibfnamefont {A.}~\bibnamefont {Gruber}}, \bibinfo
		{author} {\bibfnamefont {M.}~\bibnamefont {Domhan}}, \bibinfo {author}
		{\bibfnamefont {F.}~\bibnamefont {Jelezko}}, \ and\ \bibinfo {author}
		{\bibfnamefont {J.}~\bibnamefont {Wrachtrup}},\ }\href {\doibase
		10.1088/1367-2630/6/1/098} {\bibfield  {journal} {\bibinfo  {journal} {New J.
				Phys.}\ }\textbf {\bibinfo {volume} {6}},\ \bibinfo {pages} {98} (\bibinfo
		{year} {2004})}\BibitemShut {NoStop}%
	\bibitem [{\citenamefont {Wang}\ \emph {et~al.}(2006)\citenamefont {Wang},
		\citenamefont {Kurtsiefer}, \citenamefont {Weinfurter},\ and\ \citenamefont
		{Burchard}}]{Wang2006}%
	\BibitemOpen
	\bibfield  {author} {\bibinfo {author} {\bibfnamefont {C.}~\bibnamefont
			{Wang}}, \bibinfo {author} {\bibfnamefont {C.}~\bibnamefont {Kurtsiefer}},
		\bibinfo {author} {\bibfnamefont {H.}~\bibnamefont {Weinfurter}}, \ and\
		\bibinfo {author} {\bibfnamefont {B.}~\bibnamefont {Burchard}},\ }\href
	{\doibase 10.1088/0953-4075/39/1/005} {\bibfield  {journal} {\bibinfo
			{journal} {J. Phys. B At. Mol. Opt. Phys.}\ }\textbf {\bibinfo {volume}
			{39}},\ \bibinfo {pages} {37} (\bibinfo {year} {2006})}\BibitemShut {NoStop}%
	\bibitem [{\citenamefont {H{\"{o}}gele}\ \emph {et~al.}(2008)\citenamefont
		{H{\"{o}}gele}, \citenamefont {Galland}, \citenamefont {Winger},\ and\
		\citenamefont {{\. I}mamo{\u g}lu}}]{Hogele2008}%
	\BibitemOpen
	\bibfield  {author} {\bibinfo {author} {\bibfnamefont {A.}~\bibnamefont
			{H{\"{o}}gele}}, \bibinfo {author} {\bibfnamefont {C.}~\bibnamefont
			{Galland}}, \bibinfo {author} {\bibfnamefont {M.}~\bibnamefont {Winger}}, \
		and\ \bibinfo {author} {\bibfnamefont {A.}~\bibnamefont {{\. I}mamo{\u
					g}lu}},\ }\href {\doibase 10.1103/PhysRevLett.100.217401} {\bibfield
		{journal} {\bibinfo  {journal} {Phys. Rev. Lett.}\ }\textbf {\bibinfo
			{volume} {100}},\ \bibinfo {pages} {217401} (\bibinfo {year}
		{2008})}\BibitemShut {NoStop}%
	\bibitem [{\citenamefont {Babinec}\ \emph {et~al.}(2010)\citenamefont
		{Babinec}, \citenamefont {Hausmann}, \citenamefont {Khan}, \citenamefont
		{Zhang}, \citenamefont {Maze}, \citenamefont {Hemmer},\ and\ \citenamefont
		{Lon{\v{c}}ar}}]{Babinec2010}%
	\BibitemOpen
	\bibfield  {author} {\bibinfo {author} {\bibfnamefont {T.~M.}\ \bibnamefont
			{Babinec}}, \bibinfo {author} {\bibfnamefont {B.~J.~M.}\ \bibnamefont
			{Hausmann}}, \bibinfo {author} {\bibfnamefont {M.}~\bibnamefont {Khan}},
		\bibinfo {author} {\bibfnamefont {Y.}~\bibnamefont {Zhang}}, \bibinfo
		{author} {\bibfnamefont {J.~R.}\ \bibnamefont {Maze}}, \bibinfo {author}
		{\bibfnamefont {P.~R.}\ \bibnamefont {Hemmer}}, \ and\ \bibinfo {author}
		{\bibfnamefont {M.}~\bibnamefont {Lon{\v{c}}ar}},\ }\href {\doibase
		10.1038/nnano.2010.6} {\bibfield  {journal} {\bibinfo  {journal} {Nat.
				Nanotechnol.}\ }\textbf {\bibinfo {volume} {5}},\ \bibinfo {pages} {195}
		(\bibinfo {year} {2010})}\BibitemShut {NoStop}%
	\bibitem [{\citenamefont {He}\ \emph {et~al.}(2013)\citenamefont {He},
		\citenamefont {He}, \citenamefont {Wei}, \citenamefont {Wu}, \citenamefont
		{Atat{\"{u}}re}, \citenamefont {Schneider}, \citenamefont {H{\"{o}}fling},
		\citenamefont {Kamp}, \citenamefont {Lu},\ and\ \citenamefont
		{Pan}}]{He2013}%
	\BibitemOpen
	\bibfield  {author} {\bibinfo {author} {\bibfnamefont {Y.-M.}\ \bibnamefont
			{He}}, \bibinfo {author} {\bibfnamefont {Y.}~\bibnamefont {He}}, \bibinfo
		{author} {\bibfnamefont {Y.-J.}\ \bibnamefont {Wei}}, \bibinfo {author}
		{\bibfnamefont {D.}~\bibnamefont {Wu}}, \bibinfo {author} {\bibfnamefont
			{M.}~\bibnamefont {Atat{\"{u}}re}}, \bibinfo {author} {\bibfnamefont
			{C.}~\bibnamefont {Schneider}}, \bibinfo {author} {\bibfnamefont
			{S.}~\bibnamefont {H{\"{o}}fling}}, \bibinfo {author} {\bibfnamefont
			{M.}~\bibnamefont {Kamp}}, \bibinfo {author} {\bibfnamefont {C.-Y.}\
			\bibnamefont {Lu}}, \ and\ \bibinfo {author} {\bibfnamefont {J.-W.}\
			\bibnamefont {Pan}},\ }\href {\doibase 10.1038/nnano.2012.262} {\bibfield
		{journal} {\bibinfo  {journal} {Nat. Nanotechnol.}\ }\textbf {\bibinfo
			{volume} {8}},\ \bibinfo {pages} {213} (\bibinfo {year} {2013})}\BibitemShut
	{NoStop}%
	\bibitem [{\citenamefont {Somaschi}\ \emph {et~al.}(2016)\citenamefont
		{Somaschi}, \citenamefont {Giesz}, \citenamefont {{De Santis}}, \citenamefont
		{Loredo}, \citenamefont {Almeida}, \citenamefont {Hornecker}, \citenamefont
		{Portalupi}, \citenamefont {Grange}, \citenamefont {Ant{\'{o}}n},
		\citenamefont {Demory}, \citenamefont {G{\'{o}}mez}, \citenamefont {Sagnes},
		\citenamefont {Lanzillotti-Kimura}, \citenamefont {Lema{\'{i}}tre},
		\citenamefont {Auffeves}, \citenamefont {White}, \citenamefont {Lanco},\ and\
		\citenamefont {Senellart}}]{Somaschi2016}%
	\BibitemOpen
	\bibfield  {author} {\bibinfo {author} {\bibfnamefont {N.}~\bibnamefont
			{Somaschi}}, \bibinfo {author} {\bibfnamefont {V.}~\bibnamefont {Giesz}},
		\bibinfo {author} {\bibfnamefont {L.}~\bibnamefont {{De Santis}}}, \bibinfo
		{author} {\bibfnamefont {J.~C.}\ \bibnamefont {Loredo}}, \bibinfo {author}
		{\bibfnamefont {M.~P.}\ \bibnamefont {Almeida}}, \bibinfo {author}
		{\bibfnamefont {G.}~\bibnamefont {Hornecker}}, \bibinfo {author}
		{\bibfnamefont {S.~L.}\ \bibnamefont {Portalupi}}, \bibinfo {author}
		{\bibfnamefont {T.}~\bibnamefont {Grange}}, \bibinfo {author} {\bibfnamefont
			{C.}~\bibnamefont {Ant{\'{o}}n}}, \bibinfo {author} {\bibfnamefont
			{J.}~\bibnamefont {Demory}}, \bibinfo {author} {\bibfnamefont
			{C.}~\bibnamefont {G{\'{o}}mez}}, \bibinfo {author} {\bibfnamefont
			{I.}~\bibnamefont {Sagnes}}, \bibinfo {author} {\bibfnamefont {N.~D.}\
			\bibnamefont {Lanzillotti-Kimura}}, \bibinfo {author} {\bibfnamefont
			{A.}~\bibnamefont {Lema{\'{i}}tre}}, \bibinfo {author} {\bibfnamefont
			{A.}~\bibnamefont {Auffeves}}, \bibinfo {author} {\bibfnamefont {A.~G.}\
			\bibnamefont {White}}, \bibinfo {author} {\bibfnamefont {L.}~\bibnamefont
			{Lanco}}, \ and\ \bibinfo {author} {\bibfnamefont {P.}~\bibnamefont
			{Senellart}},\ }\href {\doibase 10.1038/nphoton.2016.23} {\bibfield
		{journal} {\bibinfo  {journal} {Nat. Photonics}\ }\textbf {\bibinfo {volume}
			{10}},\ \bibinfo {pages} {340} (\bibinfo {year} {2016})}\BibitemShut
	{NoStop}%
	\bibitem [{\citenamefont {Kuhlmann}\ \emph {et~al.}(2015)\citenamefont
		{Kuhlmann}, \citenamefont {Prechtel}, \citenamefont {Houel}, \citenamefont
		{Ludwig}, \citenamefont {Reuter}, \citenamefont {Wieck},\ and\ \citenamefont
		{Warburton}}]{Kuhlmann2015}%
	\BibitemOpen
	\bibfield  {author} {\bibinfo {author} {\bibfnamefont {A.~V.}\ \bibnamefont
			{Kuhlmann}}, \bibinfo {author} {\bibfnamefont {J.~H.}\ \bibnamefont
			{Prechtel}}, \bibinfo {author} {\bibfnamefont {J.}~\bibnamefont {Houel}},
		\bibinfo {author} {\bibfnamefont {A.}~\bibnamefont {Ludwig}}, \bibinfo
		{author} {\bibfnamefont {D.}~\bibnamefont {Reuter}}, \bibinfo {author}
		{\bibfnamefont {A.~D.}\ \bibnamefont {Wieck}}, \ and\ \bibinfo {author}
		{\bibfnamefont {R.~J.}\ \bibnamefont {Warburton}},\ }\href {\doibase
		10.1038/ncomms9204} {\bibfield  {journal} {\bibinfo  {journal} {Nat.
				Commun.}\ }\textbf {\bibinfo {volume} {6}},\ \bibinfo {pages} {8204}
		(\bibinfo {year} {2015})}\BibitemShut {NoStop}%
	\bibitem [{\citenamefont {de~Riedmatten}\ \emph {et~al.}(2003)\citenamefont
		{de~Riedmatten}, \citenamefont {Marcikic}, \citenamefont {Tittel},
		\citenamefont {Zbinden},\ and\ \citenamefont {Gisin}}]{Riedmatten2003}%
	\BibitemOpen
	\bibfield  {author} {\bibinfo {author} {\bibfnamefont {H.}~\bibnamefont
			{de~Riedmatten}}, \bibinfo {author} {\bibfnamefont {I.}~\bibnamefont
			{Marcikic}}, \bibinfo {author} {\bibfnamefont {W.}~\bibnamefont {Tittel}},
		\bibinfo {author} {\bibfnamefont {H.}~\bibnamefont {Zbinden}}, \ and\
		\bibinfo {author} {\bibfnamefont {N.}~\bibnamefont {Gisin}},\ }\href
	{\doibase 10.1103/PhysRevA.67.022301} {\bibfield  {journal} {\bibinfo
			{journal} {Phys. Rev. A}\ }\textbf {\bibinfo {volume} {67}},\ \bibinfo
		{pages} {022301} (\bibinfo {year} {2003})}\BibitemShut {NoStop}%
	\bibitem [{\citenamefont {Beugnon}\ \emph {et~al.}(2006)\citenamefont
		{Beugnon}, \citenamefont {Jones}, \citenamefont {Dingjan}, \citenamefont
		{Darqui{\'{e}}}, \citenamefont {Messin}, \citenamefont {Browaeys},\ and\
		\citenamefont {Grangier}}]{Beugnon2006}%
	\BibitemOpen
	\bibfield  {author} {\bibinfo {author} {\bibfnamefont {J.}~\bibnamefont
			{Beugnon}}, \bibinfo {author} {\bibfnamefont {M.~P.~A.}\ \bibnamefont
			{Jones}}, \bibinfo {author} {\bibfnamefont {J.}~\bibnamefont {Dingjan}},
		\bibinfo {author} {\bibfnamefont {B.}~\bibnamefont {Darqui{\'{e}}}}, \bibinfo
		{author} {\bibfnamefont {G.}~\bibnamefont {Messin}}, \bibinfo {author}
		{\bibfnamefont {A.}~\bibnamefont {Browaeys}}, \ and\ \bibinfo {author}
		{\bibfnamefont {P.}~\bibnamefont {Grangier}},\ }\href {\doibase
		10.1038/nature04628} {\bibfield  {journal} {\bibinfo  {journal} {Nature}\
		}\textbf {\bibinfo {volume} {440}},\ \bibinfo {pages} {779} (\bibinfo {year}
		{2006})}\BibitemShut {NoStop}%
	\bibitem [{\citenamefont {Maunz}\ \emph {et~al.}(2007)\citenamefont {Maunz},
		\citenamefont {Moehring}, \citenamefont {Olmschenk}, \citenamefont {Younge},
		\citenamefont {Matsukevich},\ and\ \citenamefont {Monroe}}]{Maunz2007}%
	\BibitemOpen
	\bibfield  {author} {\bibinfo {author} {\bibfnamefont {P.}~\bibnamefont
			{Maunz}}, \bibinfo {author} {\bibfnamefont {D.~L.}\ \bibnamefont {Moehring}},
		\bibinfo {author} {\bibfnamefont {S.}~\bibnamefont {Olmschenk}}, \bibinfo
		{author} {\bibfnamefont {K.~C.}\ \bibnamefont {Younge}}, \bibinfo {author}
		{\bibfnamefont {D.~N.}\ \bibnamefont {Matsukevich}}, \ and\ \bibinfo {author}
		{\bibfnamefont {C.}~\bibnamefont {Monroe}},\ }\href {\doibase
		10.1038/nphys644} {\bibfield  {journal} {\bibinfo  {journal} {Nat. Phys.}\
		}\textbf {\bibinfo {volume} {3}},\ \bibinfo {pages} {538} (\bibinfo {year}
		{2007})}\BibitemShut {NoStop}%
	\bibitem [{\citenamefont {Patel}\ \emph {et~al.}(2010)\citenamefont {Patel},
		\citenamefont {Bennett}, \citenamefont {Farrer}, \citenamefont {Nicoll},
		\citenamefont {Ritchie},\ and\ \citenamefont {Shields}}]{Patel2010}%
	\BibitemOpen
	\bibfield  {author} {\bibinfo {author} {\bibfnamefont {R.~B.}\ \bibnamefont
			{Patel}}, \bibinfo {author} {\bibfnamefont {A.~J.}\ \bibnamefont {Bennett}},
		\bibinfo {author} {\bibfnamefont {I.}~\bibnamefont {Farrer}}, \bibinfo
		{author} {\bibfnamefont {C.~A.}\ \bibnamefont {Nicoll}}, \bibinfo {author}
		{\bibfnamefont {D.~A.}\ \bibnamefont {Ritchie}}, \ and\ \bibinfo {author}
		{\bibfnamefont {A.~J.}\ \bibnamefont {Shields}},\ }\href {\doibase
		10.1038/nphoton.2010.161} {\bibfield  {journal} {\bibinfo  {journal} {Nat.
				Photonics}\ }\textbf {\bibinfo {volume} {4}},\ \bibinfo {pages} {632}
		(\bibinfo {year} {2010})}\BibitemShut {NoStop}%
	\bibitem [{\citenamefont {Lettow}\ \emph {et~al.}(2010)\citenamefont {Lettow},
		\citenamefont {Rezus}, \citenamefont {Renn}, \citenamefont {Zumofen},
		\citenamefont {Ikonen}, \citenamefont {G{\"{o}}tzinger},\ and\ \citenamefont
		{Sandoghdar}}]{Lettow2010}%
	\BibitemOpen
	\bibfield  {author} {\bibinfo {author} {\bibfnamefont {R.}~\bibnamefont
			{Lettow}}, \bibinfo {author} {\bibfnamefont {Y.~L.~A.}\ \bibnamefont
			{Rezus}}, \bibinfo {author} {\bibfnamefont {A.}~\bibnamefont {Renn}},
		\bibinfo {author} {\bibfnamefont {G.}~\bibnamefont {Zumofen}}, \bibinfo
		{author} {\bibfnamefont {E.}~\bibnamefont {Ikonen}}, \bibinfo {author}
		{\bibfnamefont {S.}~\bibnamefont {G{\"{o}}tzinger}}, \ and\ \bibinfo {author}
		{\bibfnamefont {V.}~\bibnamefont {Sandoghdar}},\ }\href {\doibase
		10.1103/PhysRevLett.104.123605} {\bibfield  {journal} {\bibinfo  {journal}
			{Phys. Rev. Lett.}\ }\textbf {\bibinfo {volume} {104}},\ \bibinfo {pages}
		{123605} (\bibinfo {year} {2010})}\BibitemShut {NoStop}%
	\bibitem [{\citenamefont {Bernien}\ \emph {et~al.}(2012)\citenamefont
		{Bernien}, \citenamefont {Childress}, \citenamefont {Robledo}, \citenamefont
		{Markham}, \citenamefont {Twitchen},\ and\ \citenamefont
		{Hanson}}]{Bernien2012}%
	\BibitemOpen
	\bibfield  {author} {\bibinfo {author} {\bibfnamefont {H.}~\bibnamefont
			{Bernien}}, \bibinfo {author} {\bibfnamefont {L.}~\bibnamefont {Childress}},
		\bibinfo {author} {\bibfnamefont {L.}~\bibnamefont {Robledo}}, \bibinfo
		{author} {\bibfnamefont {M.}~\bibnamefont {Markham}}, \bibinfo {author}
		{\bibfnamefont {D.}~\bibnamefont {Twitchen}}, \ and\ \bibinfo {author}
		{\bibfnamefont {R.}~\bibnamefont {Hanson}},\ }\href {\doibase
		10.1103/PhysRevLett.108.043604} {\bibfield  {journal} {\bibinfo  {journal}
			{Phys. Rev. Lett.}\ }\textbf {\bibinfo {volume} {108}},\ \bibinfo {pages}
		{043604} (\bibinfo {year} {2012})}\BibitemShut {NoStop}%
	\bibitem [{\citenamefont {Legero}\ \emph {et~al.}(2003)\citenamefont {Legero},
		\citenamefont {Wilk}, \citenamefont {Kuhn},\ and\ \citenamefont
		{Rempe}}]{Legero2003}%
	\BibitemOpen
	\bibfield  {author} {\bibinfo {author} {\bibfnamefont {T.}~\bibnamefont
			{Legero}}, \bibinfo {author} {\bibfnamefont {T.}~\bibnamefont {Wilk}},
		\bibinfo {author} {\bibfnamefont {A.}~\bibnamefont {Kuhn}}, \ and\ \bibinfo
		{author} {\bibfnamefont {G.}~\bibnamefont {Rempe}},\ }\href {\doibase
		10.1007/s00340-003-1337-x} {\bibfield  {journal} {\bibinfo  {journal} {Appl.
				Phys. B}\ }\textbf {\bibinfo {volume} {77}},\ \bibinfo {pages} {797}
		(\bibinfo {year} {2003})}\BibitemShut {NoStop}%
	\bibitem [{\citenamefont {Gold}\ \emph {et~al.}(2014)\citenamefont {Gold},
		\citenamefont {Thoma}, \citenamefont {Maier}, \citenamefont {Reitzenstein},
		\citenamefont {Schneider}, \citenamefont {H{\"{o}}fling},\ and\ \citenamefont
		{Kamp}}]{Gold2014}%
	\BibitemOpen
	\bibfield  {author} {\bibinfo {author} {\bibfnamefont {P.}~\bibnamefont
			{Gold}}, \bibinfo {author} {\bibfnamefont {A.}~\bibnamefont {Thoma}},
		\bibinfo {author} {\bibfnamefont {S.}~\bibnamefont {Maier}}, \bibinfo
		{author} {\bibfnamefont {S.}~\bibnamefont {Reitzenstein}}, \bibinfo {author}
		{\bibfnamefont {C.}~\bibnamefont {Schneider}}, \bibinfo {author}
		{\bibfnamefont {S.}~\bibnamefont {H{\"{o}}fling}}, \ and\ \bibinfo {author}
		{\bibfnamefont {M.}~\bibnamefont {Kamp}},\ }\href {\doibase
		10.1103/PhysRevB.89.035313} {\bibfield  {journal} {\bibinfo  {journal} {Phys.
				Rev. B}\ }\textbf {\bibinfo {volume} {89}},\ \bibinfo {pages} {035313}
		(\bibinfo {year} {2014})}\BibitemShut {NoStop}%
	\bibitem [{\citenamefont {Wang}\ \emph {et~al.}(2016)\citenamefont {Wang},
		\citenamefont {Duan}, \citenamefont {Li}, \citenamefont {Chen}, \citenamefont
		{Li}, \citenamefont {He}, \citenamefont {Chen}, \citenamefont {He},
		\citenamefont {Ding}, \citenamefont {Peng}, \citenamefont {Schneider},
		\citenamefont {Kamp}, \citenamefont {H{\"{o}}fling}, \citenamefont {Lu},\
		and\ \citenamefont {Pan}}]{Wang2016}%
	\BibitemOpen
	\bibfield  {author} {\bibinfo {author} {\bibfnamefont {H.}~\bibnamefont
			{Wang}}, \bibinfo {author} {\bibfnamefont {Z.-C.}\ \bibnamefont {Duan}},
		\bibinfo {author} {\bibfnamefont {Y.-H.}\ \bibnamefont {Li}}, \bibinfo
		{author} {\bibfnamefont {S.}~\bibnamefont {Chen}}, \bibinfo {author}
		{\bibfnamefont {J.-P.}\ \bibnamefont {Li}}, \bibinfo {author} {\bibfnamefont
			{Y.-M.}\ \bibnamefont {He}}, \bibinfo {author} {\bibfnamefont {M.-C.}\
			\bibnamefont {Chen}}, \bibinfo {author} {\bibfnamefont {Y.}~\bibnamefont
			{He}}, \bibinfo {author} {\bibfnamefont {X.}~\bibnamefont {Ding}}, \bibinfo
		{author} {\bibfnamefont {C.-Z.}\ \bibnamefont {Peng}}, \bibinfo {author}
		{\bibfnamefont {C.}~\bibnamefont {Schneider}}, \bibinfo {author}
		{\bibfnamefont {M.}~\bibnamefont {Kamp}}, \bibinfo {author} {\bibfnamefont
			{S.}~\bibnamefont {H{\"{o}}fling}}, \bibinfo {author} {\bibfnamefont {C.-Y.}\
			\bibnamefont {Lu}}, \ and\ \bibinfo {author} {\bibfnamefont {J.-W.}\
			\bibnamefont {Pan}},\ }\href {\doibase 10.1103/PhysRevLett.116.213601}
	{\bibfield  {journal} {\bibinfo  {journal} {Phys. Rev. Lett.}\ }\textbf
		{\bibinfo {volume} {116}},\ \bibinfo {pages} {213601} (\bibinfo {year}
		{2016})}\BibitemShut {NoStop}%
	\bibitem [{\citenamefont {Xie}\ and\ \citenamefont {Dunn}(1994)}]{Xie1994}%
	\BibitemOpen
	\bibfield  {author} {\bibinfo {author} {\bibfnamefont {X.~S.}\ \bibnamefont
			{Xie}}\ and\ \bibinfo {author} {\bibfnamefont {R.~C.}\ \bibnamefont {Dunn}},\
	}\href {\doibase 10.1126/science.265.5170.361} {\bibfield  {journal}
		{\bibinfo  {journal} {Science}\ }\textbf {\bibinfo {volume} {265}},\ \bibinfo
		{pages} {361} (\bibinfo {year} {1994})}\BibitemShut {NoStop}%
	\bibitem [{\citenamefont {Nirmal}\ \emph {et~al.}(1996)\citenamefont {Nirmal},
		\citenamefont {Dabbousi}, \citenamefont {Bawendi}, \citenamefont {Macklin},
		\citenamefont {Trautman}, \citenamefont {Harris},\ and\ \citenamefont
		{Brus}}]{Nirmal1996}%
	\BibitemOpen
	\bibfield  {author} {\bibinfo {author} {\bibfnamefont {M.}~\bibnamefont
			{Nirmal}}, \bibinfo {author} {\bibfnamefont {B.~O.}\ \bibnamefont
			{Dabbousi}}, \bibinfo {author} {\bibfnamefont {M.~G.}\ \bibnamefont
			{Bawendi}}, \bibinfo {author} {\bibfnamefont {J.~J.}\ \bibnamefont
			{Macklin}}, \bibinfo {author} {\bibfnamefont {J.~K.}\ \bibnamefont
			{Trautman}}, \bibinfo {author} {\bibfnamefont {T.~D.}\ \bibnamefont
			{Harris}}, \ and\ \bibinfo {author} {\bibfnamefont {L.~E.}\ \bibnamefont
			{Brus}},\ }\href {\doibase 10.1038/383802a0} {\bibfield  {journal} {\bibinfo
			{journal} {Nature}\ }\textbf {\bibinfo {volume} {383}},\ \bibinfo {pages}
		{802} (\bibinfo {year} {1996})}\BibitemShut {NoStop}%
	\bibitem [{\citenamefont {Mason}\ \emph {et~al.}(1998)\citenamefont {Mason},
		\citenamefont {Credo}, \citenamefont {Weston},\ and\ \citenamefont
		{Buratto}}]{Mason1998}%
	\BibitemOpen
	\bibfield  {author} {\bibinfo {author} {\bibfnamefont {M.~D.}\ \bibnamefont
			{Mason}}, \bibinfo {author} {\bibfnamefont {G.~M.}\ \bibnamefont {Credo}},
		\bibinfo {author} {\bibfnamefont {K.~D.}\ \bibnamefont {Weston}}, \ and\
		\bibinfo {author} {\bibfnamefont {S.~K.}\ \bibnamefont {Buratto}},\ }\href
	{\doibase 10.1103/PhysRevLett.80.5405} {\bibfield  {journal} {\bibinfo
			{journal} {Phys. Rev. Lett.}\ }\textbf {\bibinfo {volume} {80}},\ \bibinfo
		{pages} {5405} (\bibinfo {year} {1998})}\BibitemShut {NoStop}%
	\bibitem [{\citenamefont {Shimizu}\ \emph {et~al.}(2001)\citenamefont
		{Shimizu}, \citenamefont {Neuhauser}, \citenamefont {Leatherdale},
		\citenamefont {Empedocles}, \citenamefont {Woo},\ and\ \citenamefont
		{Bawendi}}]{Shimizu2001}%
	\BibitemOpen
	\bibfield  {author} {\bibinfo {author} {\bibfnamefont {K.~T.}\ \bibnamefont
			{Shimizu}}, \bibinfo {author} {\bibfnamefont {R.~G.}\ \bibnamefont
			{Neuhauser}}, \bibinfo {author} {\bibfnamefont {C.~A.}\ \bibnamefont
			{Leatherdale}}, \bibinfo {author} {\bibfnamefont {S.~A.}\ \bibnamefont
			{Empedocles}}, \bibinfo {author} {\bibfnamefont {W.~K.}\ \bibnamefont {Woo}},
		\ and\ \bibinfo {author} {\bibfnamefont {M.~G.}\ \bibnamefont {Bawendi}},\
	}\href {\doibase 10.1103/PhysRevB.63.205316} {\bibfield  {journal} {\bibinfo
			{journal} {Phys. Rev. B}\ }\textbf {\bibinfo {volume} {63}},\ \bibinfo
		{pages} {205316} (\bibinfo {year} {2001})}\BibitemShut {NoStop}%
	\bibitem [{\citenamefont {Reindl}\ \emph {et~al.}(2017)\citenamefont {Reindl},
		\citenamefont {J{\"{o}}ns}, \citenamefont {Huber}, \citenamefont {Schimpf},
		\citenamefont {Huo}, \citenamefont {Zwiller}, \citenamefont {Rastelli},\ and\
		\citenamefont {Trotta}}]{Reindl2017}%
	\BibitemOpen
	\bibfield  {author} {\bibinfo {author} {\bibfnamefont {M.}~\bibnamefont
			{Reindl}}, \bibinfo {author} {\bibfnamefont {K.~D.}\ \bibnamefont
			{J{\"{o}}ns}}, \bibinfo {author} {\bibfnamefont {D.}~\bibnamefont {Huber}},
		\bibinfo {author} {\bibfnamefont {C.}~\bibnamefont {Schimpf}}, \bibinfo
		{author} {\bibfnamefont {Y.}~\bibnamefont {Huo}}, \bibinfo {author}
		{\bibfnamefont {V.}~\bibnamefont {Zwiller}}, \bibinfo {author} {\bibfnamefont
			{A.}~\bibnamefont {Rastelli}}, \ and\ \bibinfo {author} {\bibfnamefont
			{R.}~\bibnamefont {Trotta}},\ }\href {\doibase 10.1021/acs.nanolett.7b00777}
	{\bibfield  {journal} {\bibinfo  {journal} {Nano Lett.}\ }\textbf {\bibinfo
			{volume} {17}},\ \bibinfo {pages} {4090} (\bibinfo {year}
		{2017})}\BibitemShut {NoStop}%
	\bibitem [{\citenamefont {J{\"{o}}ns}\ \emph {et~al.}(2017)\citenamefont
		{J{\"{o}}ns}, \citenamefont {Stensson}, \citenamefont {Reindl}, \citenamefont
		{Swillo}, \citenamefont {Huo}, \citenamefont {Zwiller}, \citenamefont
		{Rastelli}, \citenamefont {Trotta},\ and\ \citenamefont
		{Bj{\"{o}}rk}}]{Jons2017}%
	\BibitemOpen
	\bibfield  {author} {\bibinfo {author} {\bibfnamefont {K.~D.}\ \bibnamefont
			{J{\"{o}}ns}}, \bibinfo {author} {\bibfnamefont {K.}~\bibnamefont
			{Stensson}}, \bibinfo {author} {\bibfnamefont {M.}~\bibnamefont {Reindl}},
		\bibinfo {author} {\bibfnamefont {M.}~\bibnamefont {Swillo}}, \bibinfo
		{author} {\bibfnamefont {Y.}~\bibnamefont {Huo}}, \bibinfo {author}
		{\bibfnamefont {V.}~\bibnamefont {Zwiller}}, \bibinfo {author} {\bibfnamefont
			{A.}~\bibnamefont {Rastelli}}, \bibinfo {author} {\bibfnamefont
			{R.}~\bibnamefont {Trotta}}, \ and\ \bibinfo {author} {\bibfnamefont
			{G.}~\bibnamefont {Bj{\"{o}}rk}},\ }\href {\doibase
		10.1103/PhysRevB.96.075430} {\bibfield  {journal} {\bibinfo  {journal} {Phys.
				Rev. B}\ }\textbf {\bibinfo {volume} {96}},\ \bibinfo {pages} {075430}
		(\bibinfo {year} {2017})}\BibitemShut {NoStop}%
	\bibitem [{\citenamefont {Santori}\ \emph {et~al.}(2002)\citenamefont
		{Santori}, \citenamefont {Fattal}, \citenamefont {Vu{\v{c}}kovi{\'{c}}},
		\citenamefont {Solomon},\ and\ \citenamefont {Yamamoto}}]{Santori2002}%
	\BibitemOpen
	\bibfield  {author} {\bibinfo {author} {\bibfnamefont {C.}~\bibnamefont
			{Santori}}, \bibinfo {author} {\bibfnamefont {D.}~\bibnamefont {Fattal}},
		\bibinfo {author} {\bibfnamefont {J.}~\bibnamefont {Vu{\v{c}}kovi{\'{c}}}},
		\bibinfo {author} {\bibfnamefont {G.~S.}\ \bibnamefont {Solomon}}, \ and\
		\bibinfo {author} {\bibfnamefont {Y.}~\bibnamefont {Yamamoto}},\ }\href
	{\doibase 10.1038/nature01086} {\bibfield  {journal} {\bibinfo  {journal}
			{Nature}\ }\textbf {\bibinfo {volume} {419}},\ \bibinfo {pages} {594}
		(\bibinfo {year} {2002})}\BibitemShut {NoStop}%
	\bibitem [{\citenamefont {Davan{\c{c}}o}\ \emph {et~al.}(2014)\citenamefont
		{Davan{\c{c}}o}, \citenamefont {Hellberg}, \citenamefont {Ates},
		\citenamefont {Badolato},\ and\ \citenamefont {Srinivasan}}]{Davanco2014}%
	\BibitemOpen
	\bibfield  {author} {\bibinfo {author} {\bibfnamefont {M.}~\bibnamefont
			{Davan{\c{c}}o}}, \bibinfo {author} {\bibfnamefont {C.~S.}\ \bibnamefont
			{Hellberg}}, \bibinfo {author} {\bibfnamefont {S.}~\bibnamefont {Ates}},
		\bibinfo {author} {\bibfnamefont {A.}~\bibnamefont {Badolato}}, \ and\
		\bibinfo {author} {\bibfnamefont {K.}~\bibnamefont {Srinivasan}},\ }\href
	{\doibase 10.1103/PhysRevB.89.161303} {\bibfield  {journal} {\bibinfo
			{journal} {Phys. Rev. B}\ }\textbf {\bibinfo {volume} {89}},\ \bibinfo
		{pages} {161303} (\bibinfo {year} {2014})}\BibitemShut {NoStop}%
	\bibitem [{\citenamefont {Kettler}(2016)}]{Kettler2016PhD}%
	\BibitemOpen
	\bibfield  {author} {\bibinfo {author} {\bibfnamefont {J.}~\bibnamefont
			{Kettler}},\ }\emph {\bibinfo {title} {{Telecom-wavelength nonclassical light
				from single In(Ga)As quantum dots}}},\ \href@noop {} {\bibinfo {type} {Phd
			thesis}},\ \bibinfo  {school} {Universit{\"{a}}t Stuttgart} (\bibinfo {year}
	{2016})\BibitemShut {NoStop}%
	\bibitem [{\citenamefont {Vural}\ \emph {et~al.}(ress)\citenamefont {Vural},
		\citenamefont {Portalupi}, \citenamefont {Maisch}, \citenamefont {Kern},
		\citenamefont {Weber}, \citenamefont {Jetter}, \citenamefont {Wrachtrup},
		\citenamefont {L{\"{o}}w}, \citenamefont {Gerhardt},\ and\ \citenamefont
		{Michler}}]{Vural2018}%
	\BibitemOpen
	\bibfield  {author} {\bibinfo {author} {\bibfnamefont {H.}~\bibnamefont
			{Vural}}, \bibinfo {author} {\bibfnamefont {S.~L.}\ \bibnamefont
			{Portalupi}}, \bibinfo {author} {\bibfnamefont {J.}~\bibnamefont {Maisch}},
		\bibinfo {author} {\bibfnamefont {S.}~\bibnamefont {Kern}}, \bibinfo {author}
		{\bibfnamefont {J.~H.}\ \bibnamefont {Weber}}, \bibinfo {author}
		{\bibfnamefont {M.}~\bibnamefont {Jetter}}, \bibinfo {author} {\bibfnamefont
			{J.}~\bibnamefont {Wrachtrup}}, \bibinfo {author} {\bibfnamefont
			{R.}~\bibnamefont {L{\"{o}}w}}, \bibinfo {author} {\bibfnamefont
			{I.}~\bibnamefont {Gerhardt}}, \ and\ \bibinfo {author} {\bibfnamefont
			{P.}~\bibnamefont {Michler}},\ }\href@noop {} {\  (\bibinfo {year} {2018, in
			press})}\BibitemShut {NoStop}%
	\bibitem [{\citenamefont {Portalupi}\ \emph {et~al.}(2016)\citenamefont
		{Portalupi}, \citenamefont {Widmann}, \citenamefont {Nawrath}, \citenamefont
		{Jetter}, \citenamefont {Michler}, \citenamefont {Wrachtrup},\ and\
		\citenamefont {Gerhardt}}]{Portalupi2016}%
	\BibitemOpen
	\bibfield  {author} {\bibinfo {author} {\bibfnamefont {S.~L.}\ \bibnamefont
			{Portalupi}}, \bibinfo {author} {\bibfnamefont {M.}~\bibnamefont {Widmann}},
		\bibinfo {author} {\bibfnamefont {C.}~\bibnamefont {Nawrath}}, \bibinfo
		{author} {\bibfnamefont {M.}~\bibnamefont {Jetter}}, \bibinfo {author}
		{\bibfnamefont {P.}~\bibnamefont {Michler}}, \bibinfo {author} {\bibfnamefont
			{J.}~\bibnamefont {Wrachtrup}}, \ and\ \bibinfo {author} {\bibfnamefont
			{I.}~\bibnamefont {Gerhardt}},\ }\href {\doibase 10.1038/ncomms13632}
	{\bibfield  {journal} {\bibinfo  {journal} {Nat. Commun.}\ }\textbf {\bibinfo
			{volume} {7}},\ \bibinfo {pages} {13632} (\bibinfo {year}
		{2016})}\BibitemShut {NoStop}%
	\bibitem [{\citenamefont {Robinson}\ and\ \citenamefont
		{Goldberg}(2000)}]{Robinson2000}%
	\BibitemOpen
	\bibfield  {author} {\bibinfo {author} {\bibfnamefont {H.~D.}\ \bibnamefont
			{Robinson}}\ and\ \bibinfo {author} {\bibfnamefont {B.~B.}\ \bibnamefont
			{Goldberg}},\ }\href {\doibase 10.1103/PhysRevB.61.R5086} {\bibfield
		{journal} {\bibinfo  {journal} {Phys. Rev. B}\ }\textbf {\bibinfo {volume}
			{61}},\ \bibinfo {pages} {R5086} (\bibinfo {year} {2000})}\BibitemShut
	{NoStop}%
	\bibitem [{\citenamefont {Giesz}\ \emph {et~al.}(2015)\citenamefont {Giesz},
		\citenamefont {Portalupi}, \citenamefont {Grange}, \citenamefont
		{Ant{\'{o}}n}, \citenamefont {{De Santis}}, \citenamefont {Demory},
		\citenamefont {Somaschi}, \citenamefont {Sagnes}, \citenamefont
		{Lema{\^{i}}tre}, \citenamefont {Lanco}, \citenamefont {Auff{\`{e}}ves},\
		and\ \citenamefont {Senellart}}]{Giesz2015}%
	\BibitemOpen
	\bibfield  {author} {\bibinfo {author} {\bibfnamefont {V.}~\bibnamefont
			{Giesz}}, \bibinfo {author} {\bibfnamefont {S.~L.}\ \bibnamefont
			{Portalupi}}, \bibinfo {author} {\bibfnamefont {T.}~\bibnamefont {Grange}},
		\bibinfo {author} {\bibfnamefont {C.}~\bibnamefont {Ant{\'{o}}n}}, \bibinfo
		{author} {\bibfnamefont {L.}~\bibnamefont {{De Santis}}}, \bibinfo {author}
		{\bibfnamefont {J.}~\bibnamefont {Demory}}, \bibinfo {author} {\bibfnamefont
			{N.}~\bibnamefont {Somaschi}}, \bibinfo {author} {\bibfnamefont
			{I.}~\bibnamefont {Sagnes}}, \bibinfo {author} {\bibfnamefont
			{A.}~\bibnamefont {Lema{\^{i}}tre}}, \bibinfo {author} {\bibfnamefont
			{L.}~\bibnamefont {Lanco}}, \bibinfo {author} {\bibfnamefont
			{A.}~\bibnamefont {Auff{\`{e}}ves}}, \ and\ \bibinfo {author} {\bibfnamefont
			{P.}~\bibnamefont {Senellart}},\ }\href {\doibase 10.1103/PhysRevB.92.161302}
	{\bibfield  {journal} {\bibinfo  {journal} {Phys. Rev. B}\ }\textbf {\bibinfo
			{volume} {92}},\ \bibinfo {pages} {161302} (\bibinfo {year}
		{2015})}\BibitemShut {NoStop}%
	\bibitem [{\citenamefont {Thoma}\ \emph {et~al.}(2017)\citenamefont {Thoma},
		\citenamefont {Schnauber}, \citenamefont {B{\"{o}}hm}, \citenamefont
		{Gschrey}, \citenamefont {Schulze}, \citenamefont {Strittmatter},
		\citenamefont {Rodt}, \citenamefont {Heindel},\ and\ \citenamefont
		{Reitzenstein}}]{Thoma2017}%
	\BibitemOpen
	\bibfield  {author} {\bibinfo {author} {\bibfnamefont {A.}~\bibnamefont
			{Thoma}}, \bibinfo {author} {\bibfnamefont {P.}~\bibnamefont {Schnauber}},
		\bibinfo {author} {\bibfnamefont {J.}~\bibnamefont {B{\"{o}}hm}}, \bibinfo
		{author} {\bibfnamefont {M.}~\bibnamefont {Gschrey}}, \bibinfo {author}
		{\bibfnamefont {J.-H.}\ \bibnamefont {Schulze}}, \bibinfo {author}
		{\bibfnamefont {A.}~\bibnamefont {Strittmatter}}, \bibinfo {author}
		{\bibfnamefont {S.}~\bibnamefont {Rodt}}, \bibinfo {author} {\bibfnamefont
			{T.}~\bibnamefont {Heindel}}, \ and\ \bibinfo {author} {\bibfnamefont
			{S.}~\bibnamefont {Reitzenstein}},\ }\href {\doibase 10.1063/1.4973504}
	{\bibfield  {journal} {\bibinfo  {journal} {Appl. Phys. Lett.}\ }\textbf
		{\bibinfo {volume} {110}},\ \bibinfo {pages} {011104} (\bibinfo {year}
		{2017})}\BibitemShut {NoStop}%
	\bibitem [{\citenamefont {Zopf}\ \emph {et~al.}(2017)\citenamefont {Zopf},
		\citenamefont {Macha}, \citenamefont {Keil}, \citenamefont {Uru{\~{n}}uela},
		\citenamefont {Chen}, \citenamefont {Alt}, \citenamefont {Ratschbacher},
		\citenamefont {Ding}, \citenamefont {Meschede},\ and\ \citenamefont
		{Schmidt}}]{Zopf2017}%
	\BibitemOpen
	\bibfield  {author} {\bibinfo {author} {\bibfnamefont {M.}~\bibnamefont
			{Zopf}}, \bibinfo {author} {\bibfnamefont {T.}~\bibnamefont {Macha}},
		\bibinfo {author} {\bibfnamefont {R.}~\bibnamefont {Keil}}, \bibinfo {author}
		{\bibfnamefont {E.}~\bibnamefont {Uru{\~{n}}uela}}, \bibinfo {author}
		{\bibfnamefont {Y.}~\bibnamefont {Chen}}, \bibinfo {author} {\bibfnamefont
			{W.}~\bibnamefont {Alt}}, \bibinfo {author} {\bibfnamefont {L.}~\bibnamefont
			{Ratschbacher}}, \bibinfo {author} {\bibfnamefont {F.}~\bibnamefont {Ding}},
		\bibinfo {author} {\bibfnamefont {D.}~\bibnamefont {Meschede}}, \ and\
		\bibinfo {author} {\bibfnamefont {O.~G.}\ \bibnamefont {Schmidt}},\ }\href
	{http://arxiv.org/abs/1712.08158} {\bibfield  {journal} {\bibinfo  {journal}
			{arXiv}\ ,\ \bibinfo {pages} {1}} (\bibinfo {year} {2017})},\ \Eprint
	{http://arxiv.org/abs/1712.08158} {arXiv:1712.08158} \BibitemShut {NoStop}%
	\bibitem [{\citenamefont {Nguyen}\ \emph {et~al.}(2013)\citenamefont {Nguyen},
		\citenamefont {Sallen}, \citenamefont {Abbarchi}, \citenamefont {Ferreira},
		\citenamefont {Voisin}, \citenamefont {Roussignol}, \citenamefont
		{Cassabois},\ and\ \citenamefont {Diederichs}}]{Nguyen2013}%
	\BibitemOpen
	\bibfield  {author} {\bibinfo {author} {\bibfnamefont {H.~S.}\ \bibnamefont
			{Nguyen}}, \bibinfo {author} {\bibfnamefont {G.}~\bibnamefont {Sallen}},
		\bibinfo {author} {\bibfnamefont {M.}~\bibnamefont {Abbarchi}}, \bibinfo
		{author} {\bibfnamefont {R.}~\bibnamefont {Ferreira}}, \bibinfo {author}
		{\bibfnamefont {C.}~\bibnamefont {Voisin}}, \bibinfo {author} {\bibfnamefont
			{P.}~\bibnamefont {Roussignol}}, \bibinfo {author} {\bibfnamefont
			{G.}~\bibnamefont {Cassabois}}, \ and\ \bibinfo {author} {\bibfnamefont
			{C.}~\bibnamefont {Diederichs}},\ }\href {\doibase
		10.1103/PhysRevB.87.115305} {\bibfield  {journal} {\bibinfo  {journal} {Phys.
				Rev. B}\ }\textbf {\bibinfo {volume} {87}},\ \bibinfo {pages} {115305}
		(\bibinfo {year} {2013})}\BibitemShut {NoStop}%
\end{thebibliography}
%
\section{Supplementary information to ``Overcoming correlation fluctuations in two-photon interference experiments with differently bright and independently blinking remote quantum emitters"}

\section{Simulation}
\label{sec:simulation}
\subsection{Monte Carlo simulation of random population}
For the majority of the QDs on the sample under study, two characteristics are found to coincide: (i) the necessity to apply a weak above-band laser in order to obtain a strong resonance fluorescence signal, and (ii) the observation of a bunching signature in second-order autocorrelation histograms. The former has been conclusively modeled by describing the random population dynamics of a QD in the vicinity of a nearby carrier trap by means of coupling differential equations to the individual carrier configurations [44, main text]. While this method offers a convenient way to describe the steady state situation of a single system, the extension to the description of two emitters with additional access to individual time-resolved blinking dynamics is not straightforward.

\begin{figure}
	\centering
	\includegraphics{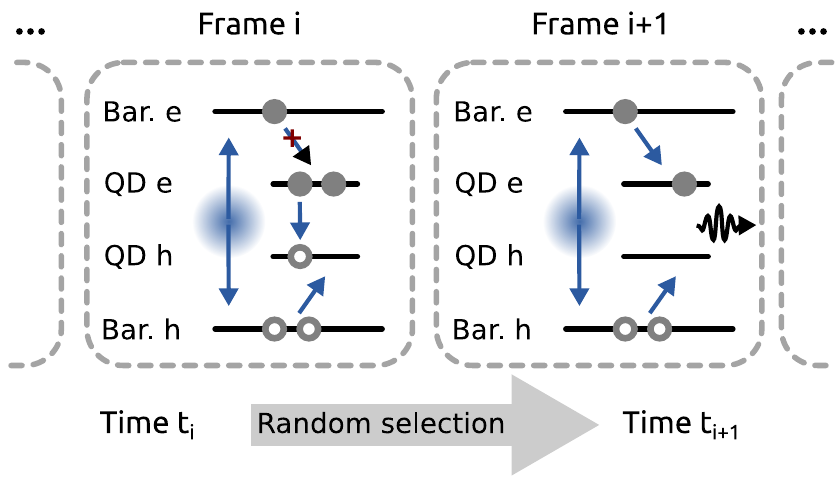}
	\caption{Two random time frames for a simple example system consisting of barrier levels and a QD s-shell with a population limit of 2. In order to simulate the evolution between subsequent frames, from all possible transitions (considering relaxation, recombination, excitation), a random realization is drawn, respecting the individual transition probabilities. In case of a recombination event, a photon time stamp is registered.\label{fig:figS1}}
\end{figure}

An alternative method can be found by using a Monte-Carlo simulation for the evolution of the microscopic carrier configurations, i.e., following individual random population and depopulation events [37, main text]. This can be seen as a discrete implementation of a rate equation model, respecting population limits for confined states: the system configuration is continuously followed on a random path obeying all involved transition probabilities. For each configuration change, a snapshot of the configuration in the respective time frame is registered. Two example time frames are displayed in Figure \ref{fig:figS1}. The transition from one frame to the next frame is obtained by drawing one random realization of all possible population and depopulation options, which compose of spontaneous recombination, relaxation and fixed initialization events. It should be noted, that this method, as for rate equation or random population models, does not offer any access to coherent population dynamics, e.g. Rabi oscillations or similar. It is, however, sufficient to conclusively capture population dynamics governed by the (typically incoherent) environment, and describe the impact on the intensity dynamics of the emitted light. It is further offering a convenient method to explore the consequences upon photon correlation histograms, as photon arrival times are directly provided, in analogy to a measurement configuration applying photon time tagging.
\begin{figure}
	\centering
	\includegraphics{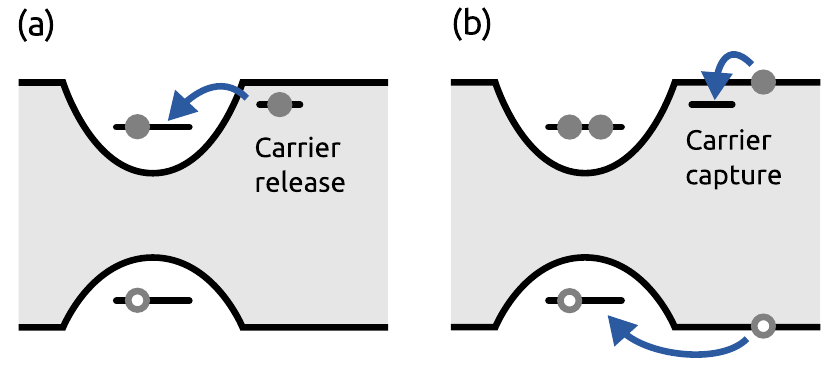}
	\caption{A simple model to describe a blinking QD transition due to a nearby charge carrier trap. Carrier release (a) and carrier capture (b) by the trap state result in the charging or neutralization of the initially uncharged QD. The time constants describing carrier capture and release processes, as well as the above-band excitation rate determine the blinking dynamics and thus the bunching behavior observed in photon correlation measurements. \label{fig:figS2}}	
\end{figure}
\subsection{Modeling of a blinking emitter}
The most simple system to sufficiently describe blinking dynamics as observed in QDR and QDB is found to consist of: (i) a barrier level for electrons and holes, (ii) a two-fold degenerate s-shell for electrons and holes, and (iii) a single charge carrier trap with a relaxation channel to the QD s-shell. The relaxation of a trapped carrier to the QD on one hand, and the trapping of one carrier from an above-band electron-hole pair on the other hand, constitute opposite charging mechanisms for the QD (see Figure \ref{fig:figS2}. This results in a randomly switching QD charge state, consequently leading to the observation of blinking in a resonantly excited QD transition.

\begin{figure}
	\centering
	\includegraphics{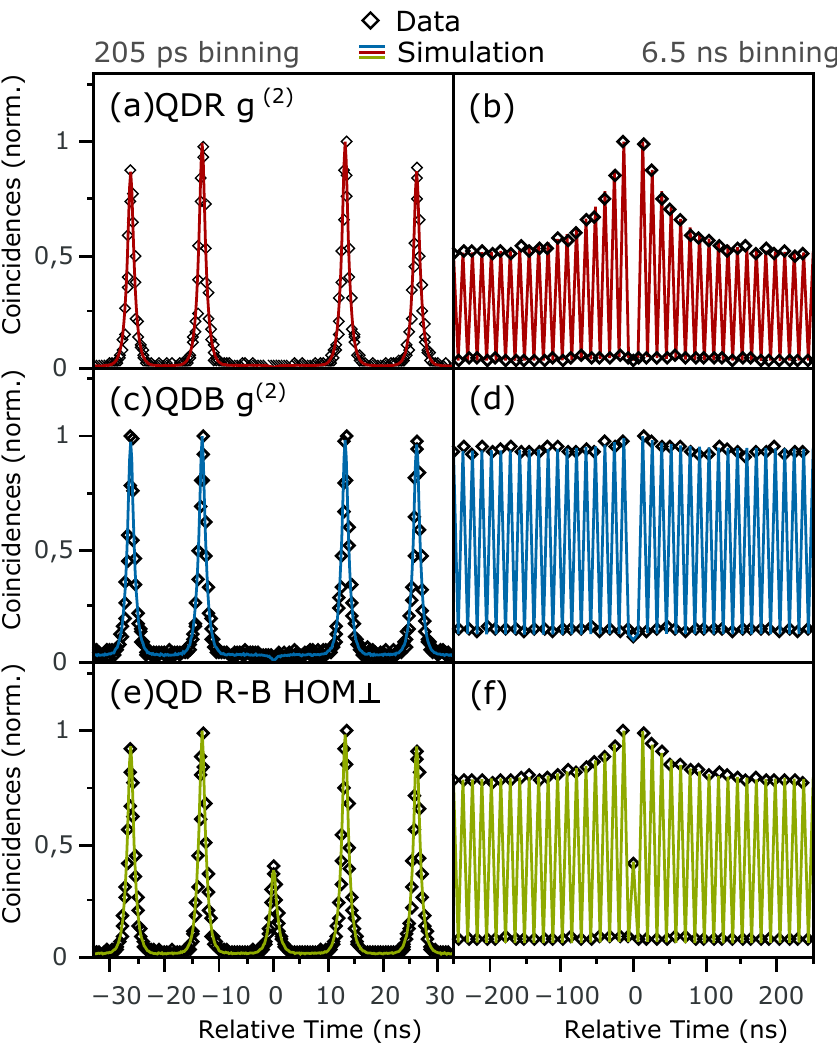}
	\caption{QD-pair \#1: (a-d) Simulated and measured intensity autocorrelation functions for QDR and QDB displayed for short and long correlation windows. Using a system as described in Fig. S2, the simulation parameters are determined by the bunching time constant, bunching strength and above-band excitation rate. Using the parameter sets extracted for both QDs, the TPI measurement for distinguishable photons from QDR and QDB (e-f) is perfectly matched by the simulation.  \label{fig:figS3}}
\end{figure}

The on- and off-periods of a distinct charge state consequently evoke a bunching signature in intensity correlation measurements, which is mainly governed by the carrier capture and release times of the trap level, as well as the creation rate for above-band electron-hole pairs. For the QDs under study, the cw above-band excitation further resulted in a non-negligible signature superimposed to the  pulsed intensity correlation histograms,thus allowing to determine all of the aforementioned transition rates. Figure \ref{fig:figS3}a-d is showing simulated and measured intensity autocorrelation histograms for QDR and QDB on a short time scale to identify the contribution of the above-band excitation and a sufficiently long time scale to identify the bunching behavior. Using the hereby defined parameter set, a simulated HOM histogram between distinguishable photons from QD A and B is perfectly matching the experimental data (Figure \ref{fig:figS3}e,f).

\section{Coincidence Mismatch}
\label{sec:coinc_mismatch}
\begin{figure}
	\centering
	\includegraphics{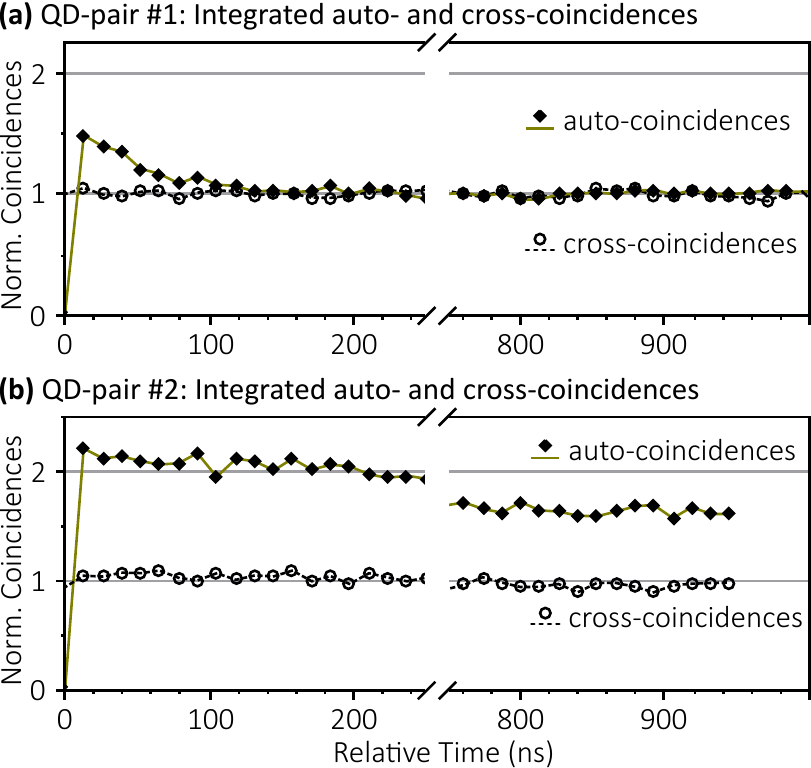}
	\caption{QD-pair \#1: Integrated auto- and cross-coincidence events per repetition extracted from Figure 2d, main article. Bunching in auto-coincidence events on the time scale of blinking. Anti-bunching of auto-coincidence can be observed at zero time delay due to the single-photon nature of the emitters. Approximation of both types of coincidences for larger time-scales, but matching only observed due to mismatch in beam splitting ratio, i.e. $\beta = 46/54$ (compare Figure \ref{fig:figS5}). QD-pair \#2: Same setup scheme as in (a), however, with much longer relevant blinking time scales. Normalization on the Poissonian level impractical.\label{fig:figS4}}
\end{figure}
\begin{figure}
	\centering
	\includegraphics{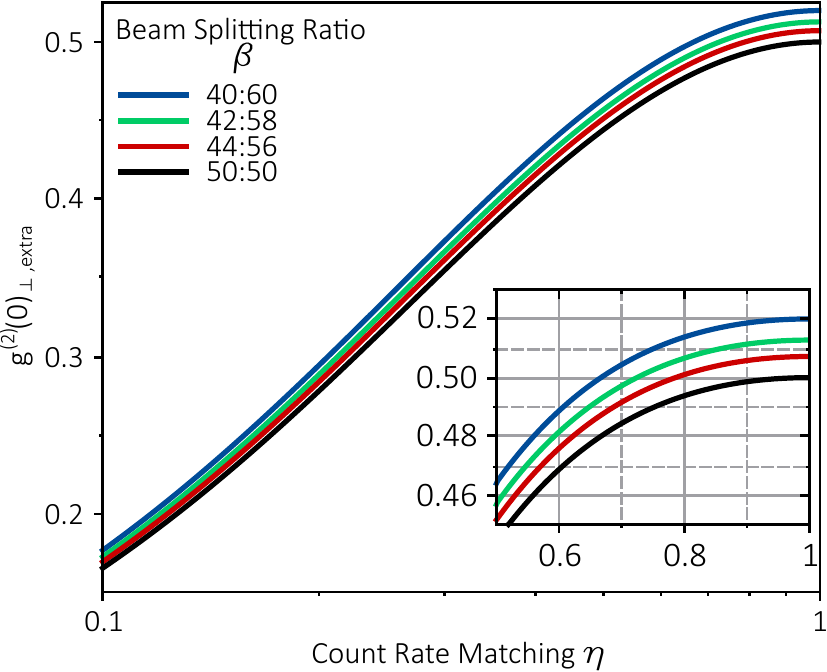}
	\caption{Plot of $\gtwoo_\text{$\perp$,extra}$, the extrapolated $\gtwoo_\text{$\perp$}$ in dependence of count rate matching $\eta$ for different beam splitter ratios $\beta$ following (\ref{eq:g2extra2}). The extrapolated value enables determination of the TPI contrast within a single measurement of $\gtwoo_\text{$\parallel$,exp}$ when $\gtwot_\text{$\parallel$,exp}$ is normalized to the Poissonian level. \label{fig:figS5}}
\end{figure}
As defined in the main article, two types of coincidences occur in a TPI experiment with remote emitters. Coincidences within the individual photon streams are referred to as auto-coincidences and coincidences between the two emitters are referred to as cross-coincidences. By means of an additional fiber the photon streams of the two emitters become unsynchronized in time. Then, coincidences between the two photon streams are separated in the resulting correlation measurement (compare main text Figure 2b). It is then possible to identify auto- and cross-coincidences making a quantification possible. Figure \ref{fig:figS4} shows the integrated coincidence counts of the respective types (extracted from main article Figure 2d). At zero time delay, antibunching of the auto-coincidences can be observed due to the single-photon nature of the two emitters. Clustering in the single-photon stream is indicated by bunching of the auto-coincidences. Cross-coincidences, however, do not exhibit any bunching or correlation - the two emitters are uncorrelated in time. By means of such a distinction, $\gtwoo_\perp$, which is obtained by normalization of the zero-delay peak ($A_0$) to coincidences accumulated at the side peaks (${A_\text{side}}$), can be expressed in terms of auto-  ($A_\text{auto}$) and cross-coincidences ($A_\text{cross}$):
\begin{equation}
	\gtwoo_\perp = \frac{A_0}{A_\text{side}} = \frac{A_\text{cross}}{A_\text{auto}+A_\text{cross}}\label{eq:g2extra1}
\end{equation}   

As previously discussed, auto-coincidences show temporal correlations such that $A_\text{auto}=A_\text{auto}(\tau)$, whereas $A_\text{cross}=const$. Taking into account blinking (compare Santori et al. (2001) [12, main text]), count rates $c_i$ of the individual photon streams and beam splitter ratio T:R with transmission $T$ and reflectivity $R=1-T$, where $T\in[0,0.5]$, $A_\text{auto}$ and $A_\text{cross}$ can be further written as:
\begin{align}
	A_\text{auto} 	&= \Bigg[\left(1+\frac{\tau_\text{1,off}}{\tau_\text{1,on}}e^{-\left(\frac{1}{\tau_\text{1,off}}+ \frac{1}{\tau_\text{1,on}}\right)|m\tau_\text{rep}|}\right)c_1^2\notag\\ &+\left(1+\frac{\tau_\text{2,off}}{\tau_\text{2,on}}e^{-\left(\frac{1}{\tau_\text{2,off}} + \frac{1}{\tau_\text{2,on}}\right) |m\tau_\text{rep}|}\right)c_2^2\Bigg] RT\\
	A_\text{cross} 	&= c_1c_2\left[R^2+T^2\right]
\end{align}
For normalization on the Poissonian level, i.e. for $A_\text{side}$ at $|m\cdot\tau_\text{rep}|\gg \tau_\text{on}$, with $m\in\mathbb{N}^+$, it is possible to predict $\gtwoo_\perp$ as we enable an extrapolation via monitoring of the individual count rates. Then (\ref{eq:g2extra1}) becomes the extrapolated $\gtwoo_\text{$\perp$}$:
\begin{equation}
	\gtwoo_\text{$\perp$,extra} = \frac{c_1c_2\left[R^2+T^2\right]}{\left[c_1^2+c_2^2\right]RT+c_1c_2\left[R^2+T^2\right]}
\end{equation}
This equation can be further simplified as we define the matching of count rates $c_{1,2}$ as $\eta\equiv c_2/c_1$ (where $c_2\leq c_1$) and matching of beam splitter ratio as $\beta \equiv T/R$ (where $T \leq R$), leading to:
\begin{equation}
	\gtwoo_\text{$\perp$,extra} = \frac{\eta\left[1+\beta^2\right]}{\eta\left[1+\beta^2\right]+\beta\left[1+\eta^2\right]},\label{eq:g2extra2}
\end{equation}
with $\eta,\beta \in[0,1]$.
Figure \ref{fig:figS5} shows the extrapolated $\gtwoo_\text{$\perp$,extra}$ in dependence of count rate matching $\eta$ for different beam splitting ratios $\beta$. For perfect beam splitting ratio, i.e $\beta = 50/50$, (\ref{eq:g2extra2}) leads to 
\begin{equation}
	\gtwoo_\text{$\perp$,extra} = \frac{2\eta}{\left(1+\eta\right)^2},
\end{equation}
which is already mentioned in the main article.

To extract $\eta$ from the measurement, the detection time trace of detector $A$ and $B$ have to be recorded. If temporally only one of the two emitters $c_i$ is unblocked, it is possible to directly extract the count rate matching at the beam splitter via the resulting detection rates $d_{A,i}$ and $d_{B,i}$:
\begin{align}
	\tilde{\eta}_A &= \frac{d_{A,2}}{d_{A,1}} = \frac{c_2\cdot\beta}{c_1/\beta}=\frac{c_2}{c_1}\cdot\beta^2\\
	\tilde{\eta}_B &= \frac{d_{B,2}}{d_{B,1}} = \frac{c_2/\beta}{c_1\cdot\beta}=\frac{c_2}{c_1}\cdot \frac{1}{\beta^2}\\
	\eta &= \sqrt{\tilde{\eta}_A \tilde{\eta}_B} = \frac{c_2}{c_1}
\end{align}
In this study, however, beam blocking was not carried out. Therefore, a drop in detection rate was directly linked to the more unstable emitter. This was then further utilized to extract $\eta$ as the initial detection trace was supposed to start with two equally bright emitters.

It is important to mention, that $\gtwoo_\text{$\perp$}$ can only be extrapolated from pure statistical consideration if the Possonian level is taken for normalization. However, blinking can in principle span over several time orders [36, main text], which may result in impractical correlation time windows as it is the case for Figure \ref{fig:figS5}b. Here the emission stems from QD/pair \#2. As one can see, the Poissonian level is not reached even within a correlation time window of \SI{10}{\micro\second}. If this is the case, the standard approach might not be feasible and the demonstrated new method should be exploited. To obtain the actual degree of indistinguishability, beam splitter correction is applied following Ref. [35, main text] including splitting ratio and mode overlap.
\begin{figure*}
	\centering
	\includegraphics{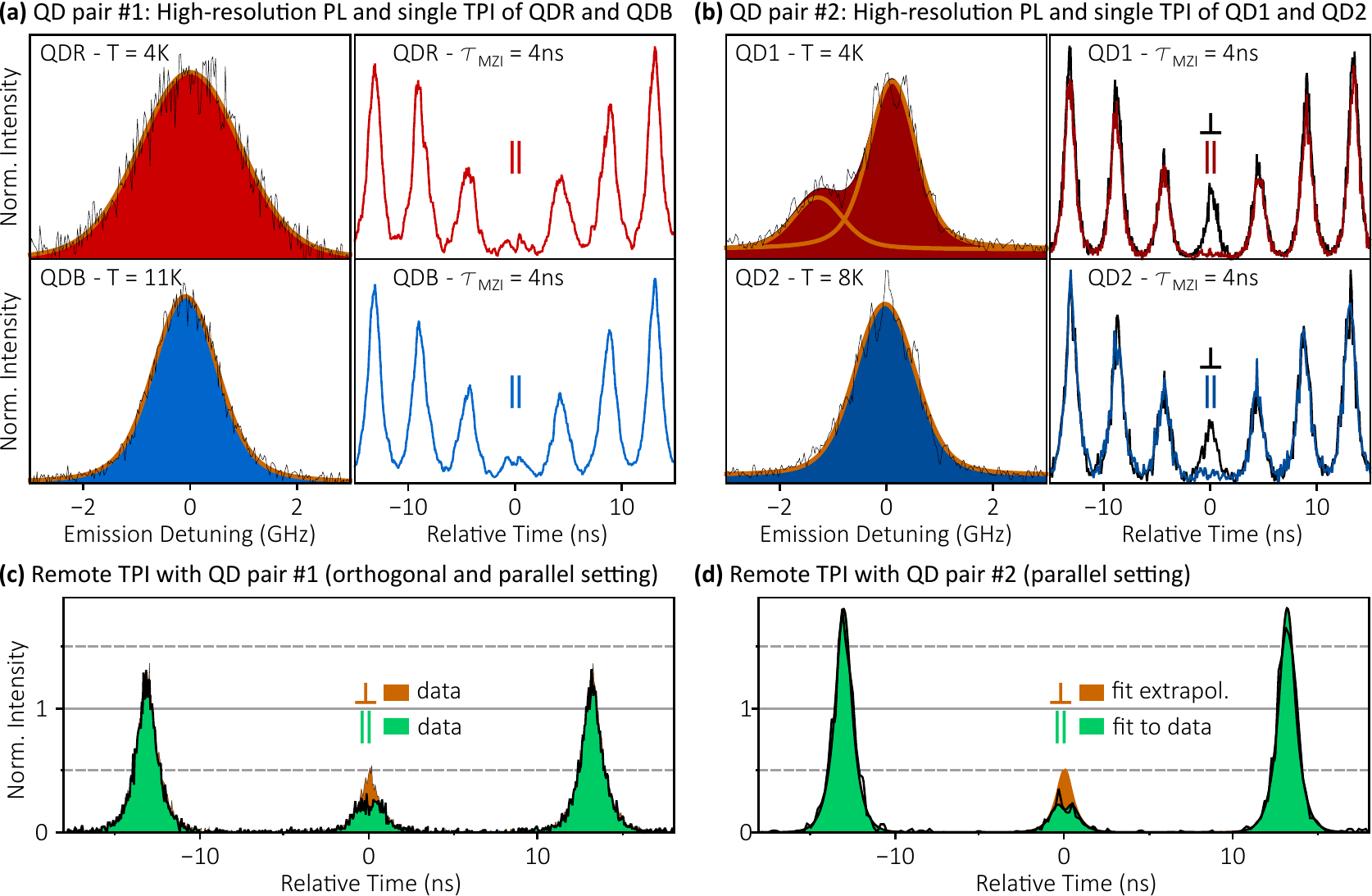}
	\caption{QD-pair \#1: (a,b) High-resolution PL of the four QDs as well as the respective TPI measurement of consecutively emitted photons from the individual emitters. (c) Remote TPI measurement of QD-pair \#1. (d) Remote TPI measurement of QD-pair \#2. Normalization on the Poissonian level is not feasible, hence, the orthogonal measurement is an extrapolation from the fit of the data neglecting the interference term of the fit function.\label{fig:figS6}}
\end{figure*}
\section{Two-photon interference with QD-pair \#1 and QD-pair \#2}
\label{sec:TPI_pair2}
In this study, two QD-pairs are compared to illustrate the influence of spectral wandering of the individual QD on the remote TPI contrast. QD-pair \#1 is grown via metalorganic vapor phase epitaxy [39, main text], whereas QD-pair \#2 is grown via molecular beam epitaxy [17, main text]. All experiments are carried out under resonant excitation via a Ti:Sapphire laser at a repetition rate of \SI{76.2}{\mega\hertz}, i.e., $\tau_\text{rep}=\SI{13.12}{\nano\second}$, and a pulse width of \SI{3}{\pico\second}. The charged exciton transitions of QDR/B and QD1/2 are individually initialized by means of coherent $\pi$-pulse excitation. In addition, continuous-wave above-band excitation was individually applied to QDR/B, as well as whitelight excitation to QD1/2. A scanning Fabry-Perot interferometer is utilized for high-resolution PL measurements. It has a free spectral range $\text{FSR}=\SI{15}{\giga\hertz}$ and a resolution of $\Delta\nu_\text{FPI}\approx\SI{0.1}{\giga\hertz}$. TPI experiments with the individual emitters are carried out via interfering consecutively emitted photons by means of an unbalanced MZI having a time delay of $\tau_\text{MZI}\approx\SI{4}{\nano\second}$. 
In Figure \ref{fig:figS6}a,b the high-resolution PL and single emitter TPI measurement is respectively shown for QD-pair \#1 (a) and \#2 (b). The extracted data (see Table I, main text) shows that inhomogeneous broadening does not necessarily reflect the degree of indistinguishability of consecutively emitted photons. Instead it is rather driven by the time constant of the spectral diffusion processes. Here, QD-pair \#1 exhibits diffusion time constant on the order of the MZI time delay. The TPI measurements with QD1 and QD2, instead, suggest slower spectral diffusion as the TPI visibility of the single emitters is much higher. 
However, in case of remote TPI, spectral diffusion time constants do not play a role as the interfering photons have no spectral correlation. This can be seen in Figure \ref{fig:figS6}c,d where the remote TPI measurements with QD-pair \#1 (c) and \#2 (d) is shown. In this case, the similarities in the inhomogeneous broadening can be observed in the TPI measurement.
\end{document}